\documentclass[journal=jacsat,manuscript=article]{achemso}

\usepackage[version=3]{mhchem} 
\usepackage[dvipsnames]{xcolor}
\usepackage{caption}
\usepackage{svg}
\usepackage{subcaption}
\usepackage{enumitem,amssymb}
\usepackage{pdflscape}
\newlist{todolist}{itemize}{2}
\setlist[todolist]{label=$\square$}

\newcommand*{\up}{\uparrow}

\newcommand*{\dn}{\downarrow}

\newcommand*{\ket}[1]{\left | {#1} \right\rangle}

\newcommand*{\eaabb}{\left | \up \up \dn \dn \right \rangle}
\newcommand*{\eabab}{\left | \up \dn \up \dn \right \rangle}
\newcommand*{\eabba}{\left | \up \dn \dn \up \right \rangle}
\newcommand*{\ebbaa}{\left | \dn \dn \up \up \right \rangle}
\newcommand*{\ebaba}{\left | \dn \up \dn \up \right \rangle}
\newcommand*{\ebaab}{\left | \dn \up \up \dn \right \rangle}
\newcommand*{\aabb}{$\left | \up \up \dn \dn \right \rangle$}
\newcommand*{\abab}{$\left | \up \dn \up \dn \right \rangle$}
\newcommand*{\abba}{$\left | \up \dn \dn \up \right \rangle$}
\newcommand*{\bbaa}{$\left | \dn \dn \up \up \right \rangle$}
\newcommand*{\baba}{$\left | \dn \up \dn \up \right \rangle$}
\newcommand*{\baab}{$\left | \dn \up \up \dn \right \rangle$}

\newcommand*{\angs}{$\text{\AA}$ }
\newcommand*\myatop[2]{\genfrac{}{}{0pt}{}{#1}{#2}}

\author{Scott M. Garner}
\affiliation{Department of Chemistry, University of California, Berkeley, CA, 94720, United States}
\alsoaffiliation{Chemical Sciences Division, Lawrence Berkeley National Laboratory, Berkeley, CA, 94720, United States}
\author{Eric A. Haugen}
\affiliation{Department of Chemistry, University of California, Berkeley, CA, 94720, United States}
\alsoaffiliation{Chemical Sciences Division, Lawrence Berkeley National Laboratory, Berkeley, CA, 94720, United States}
\author{Stephen R. Leone}
\affiliation{Department of Chemistry, University of California, Berkeley, CA, 94720, United States}
\alsoaffiliation{Chemical Sciences Division, Lawrence Berkeley National Laboratory, Berkeley, CA, 94720, United States}
\alsoaffiliation{Department of Physics, University of California, Berkeley, CA 94720, United States}
 \author{Eric Neuscamman}
\affiliation{Department of Chemistry, University of California, Berkeley, CA, 94720, United States}
\alsoaffiliation{Chemical Sciences Division, Lawrence Berkeley National Laboratory, Berkeley, CA, 94720, United States}\email{eneuscamman@berkeley.edu}

\title{Spin Coupling Effect on Geometry-Dependent X-ray Absorption of Diradicals}

\begin{document}

\begin{tocentry}
\includegraphics[width=8.25cm,height=4.45cm]{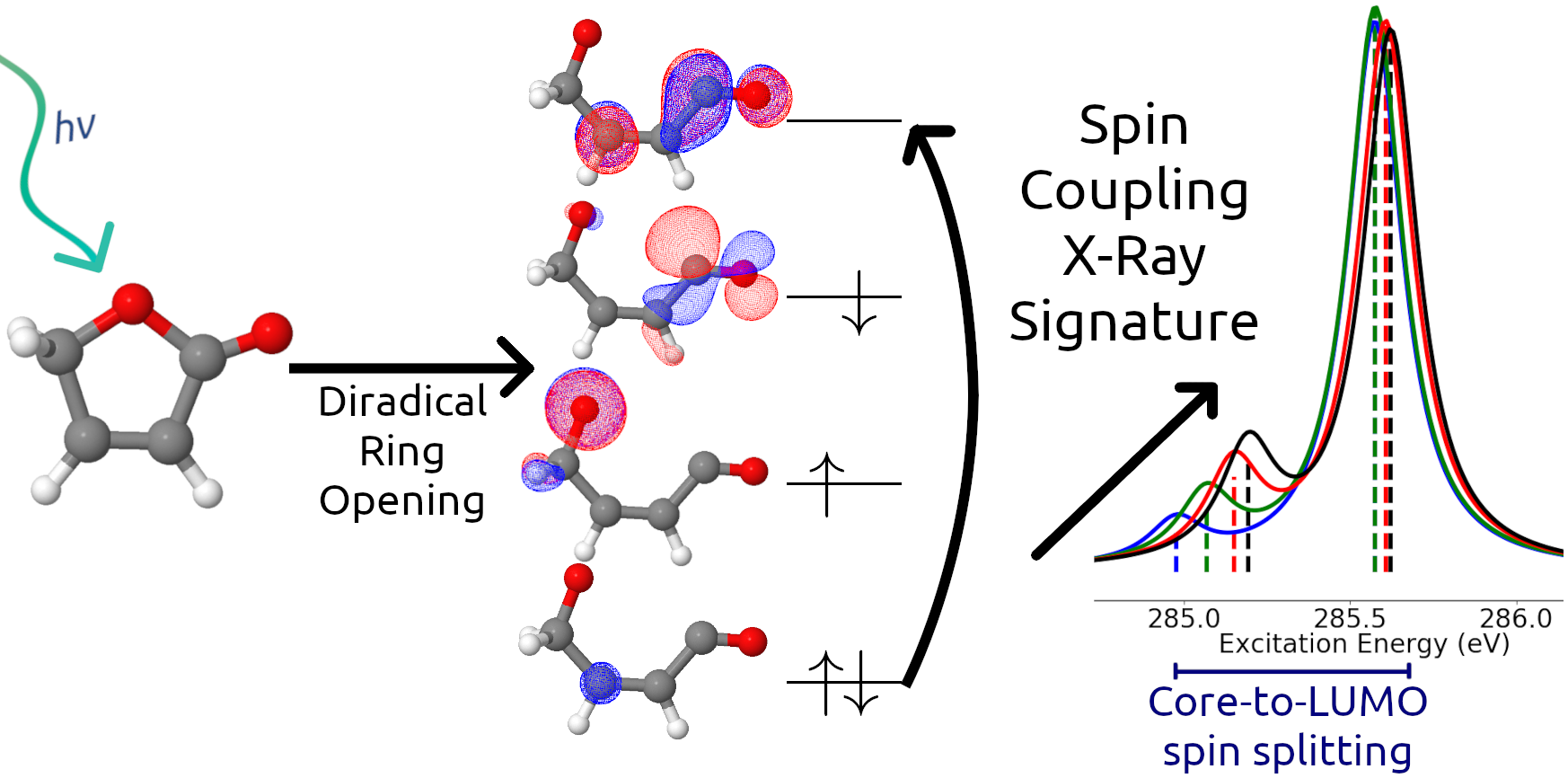}
\end{tocentry}

\begin{abstract}
We theoretically investigate the influence of diradical electron spin coupling on the time-resolved X-ray absorption spectra of the photochemical ring opening of furanone.
We predict geometry dependent carbon K-edge signals involving transitions from core orbitals to both singly and unoccupied molecular orbitals.
The most obvious features of the ring opening come from the carbon atom directly involved in the bond breaking, through its transition to both the newly formed SOMO and the available LUMO state.
In addition to this primary feature, the singlet spin coupling of four unpaired electrons that arises in the core-to-LUMO states creates additional geometry dependence in some spectral features, with both oscillator strengths and relative excitation energies varying observably as a function of the ring opening.
We attribute this behavior to a spin-occupancy-induced selection rule, which occurs when singlet spin coupling is enforced in the diradical state.
Notably, one of these geometry-sensitive core-to-LUMO transitions excites core electrons from a backbone carbon not involved in the bond breaking, providing a novel non-local X-ray probe of chemical dynamics arising from electron spin coupling.
\end{abstract}

\section{Introduction}

Time-resolved X-ray spectroscopy is revolutionizing the understanding of many chemical processes, allowing observation of electron and nuclear dynamics down to attosecond time scales\cite{kraus2018ultrafast_XASDynamics_Review,Young_2018_Roadmap,bhattacherjee2018ultrafast_XASMolecularReview}.
Functionally, X-ray light probes inner-shell core electrons, with different elements easily identifiable by their distinct spectroscopic energy edges.
Focusing on X-ray absorption (XAS), each unique atom's core to valence excitation manifold contributes its own fingerprint to the total spectrum.
Assigning transitions to individual atoms thus gives insight to the local chemical environment.
In a time-resolved experiment, the changes in the spectra as a function of time can thus be mapped onto local changes in electronic structure\cite{goulielmakis2010real_Kr_Dynamics,vura2013femtosecond_IronOxidationStateXAS,wolf2017probing_InternalConversionRevealedByOxygenKEdge}, spin state\cite{zhang2019tracking_IronSpinCrossoverXAS,bhattacherjee2017ultrafast_ISCAcetylacetoneXAS}, and chemical bonding\cite{chatterley2016dissociation_FerroceniumDissociationXUV,Pertot2017_CF4SF6BondBreakingXAS,bhattacherjee2018photoinduced_FurfuralXASOpening,toulson2023probing_GessnerThiopheneIodine}.
With sources of short-pulse, broadband X-ray light becoming increasingly available, it is beneficial to be able to predict and analyze the spectroscopic features arising from different common chemical motifs.

One motif of recent interest is radical species.
For single radical species--that is, open shell doublets with one electron residing in a singly occupied molecular orbital (SOMO)--a recent review by Carravetta \textit{et. al.}\cite{carravetta2022x_XAS_Cations_Review} divides the XAS spectra into three principal sets of features:
the first two regions are characterized by core-to-SOMO and spin-coupled core-to-valence transitions, respectively, while the third region contains densely packed core-to-Rydberg transitions.
It is the first two regions that are of interest here, ultimately for diradicals.
Core-to-SOMO excitations typically appear experimentally as the lowest energy features for radicals and ions and result in a final state with a single unpaired electron remaining in the core.
Promotion of a core electron into an unoccupied orbital--often the lowest unoccupied molecular orbital (LUMO)--results in a state with three unpaired electrons in three spatially distinct orbitals.
Coupling these electrons, or equivalently coupling three spin-$\frac{1}{2}$ systems, results in two doublet states, and therefore two possible spectroscopically accessible transitions where the initial state is an open shell doublet.
This results in spin-induced splitting of these core-to-unoccupied valence transitions, a unique feature that has been experimentally observed in a number of small radical molecules \cite{alagia2013soft_AllylRadicalXAS,couto2020carbon_COPlus_XAS_Spectra,lindblad2020x_NitrogenPlus_XAS,bari2019inner_NH0to3Plus_XAS,ridente2022femtosecond_methaneCation_XAS} as well as studied theoretically\cite{hait2020accurate_Dip_ROKSCoreRadicalRemixing,zhao2021dynamic_DynamicThenStatic,carravetta2022x_XAS_Cations_Review}.
Further, in time-resolved XAS measurements of methane and benzene cations, insights into Jahn-Teller relaxation timescales, energetics, and related vibrational dynamics have been inferred from both time-resolved core-to-SOMO (methane cation\cite{ridente2022femtosecond_methaneCation_XAS}) and spin-split core-to-LUMO (benzene cation\cite{epshtein2020table_BenzenePlus_Experiment,vidal2020interplay_BenezenPlus_Theory}) transitions.
Here, we are interested in extending the understanding of XAS predictions to spectral properties of diradicals, i.e. molecules in which two electrons singly occupy two spatially distinct orbitals.
Motivated by utilizing XAS to study ultrafast chemical dynamics, we will focus exclusively on singlet coupled diradicals, rationalizing that once a singlet diradical is created (for example, by photoexcitation), other relaxation dynamics may occur on timescales faster than typical intersystem crossing timescales\cite{mcquarrie1997physical_simonandmcquarrie}.

Upon probing a diradical with X-ray light, core-to-SOMO transitions provide a direct isolated probe of the SOMO orbitals on specific atoms.
Core-to-LUMO transitions lead to states with four unpaired electrons in four orbitals: the two diradical SOMO electrons, the partially filled core orbital electron, and the partially filled LUMO electron.
The spin coupling between these electrons makes the final spectra more interesting, challenging, and potentially useful.
Specifically, we investigate how the spin coupling between four unpaired electrons influences the spin splitting and relative X-ray oscillator strengths of the two components of a spin-coupled core-to-LUMO transition. 
Focusing on photochemical ring opening of the furanone molecule, we predict two geometry dependent carbon K-edge XAS features. 
First, a red-shift of the core-to-SOMO transition where both the core and SOMO orbitals are localized on the carbon atom involved in the bond breaking. 
Second, we predict a monotonic increase and shift in X-ray absorption intensity versus ring opening distance for one of the two components of the core-to-LUMO excitation from a spectator carbon (i.e. a carbon not involved in the bond breaking). 
We attribute this increase in intensity to changes in the details of spin coupling between the four unpaired electrons as the ring opens and the diradical electrons are spatially separated.
Of particular interest, this increase and shift in one of the spin-coupled absorption peaks can provide a secondary time-dependent observational spectroscopic probe of the ring opening dynamics coming from a carbon atom spatially distant from the bond breaking process.

\begin{figure}[htpb]
    \centering
    \includegraphics[width=\textwidth]{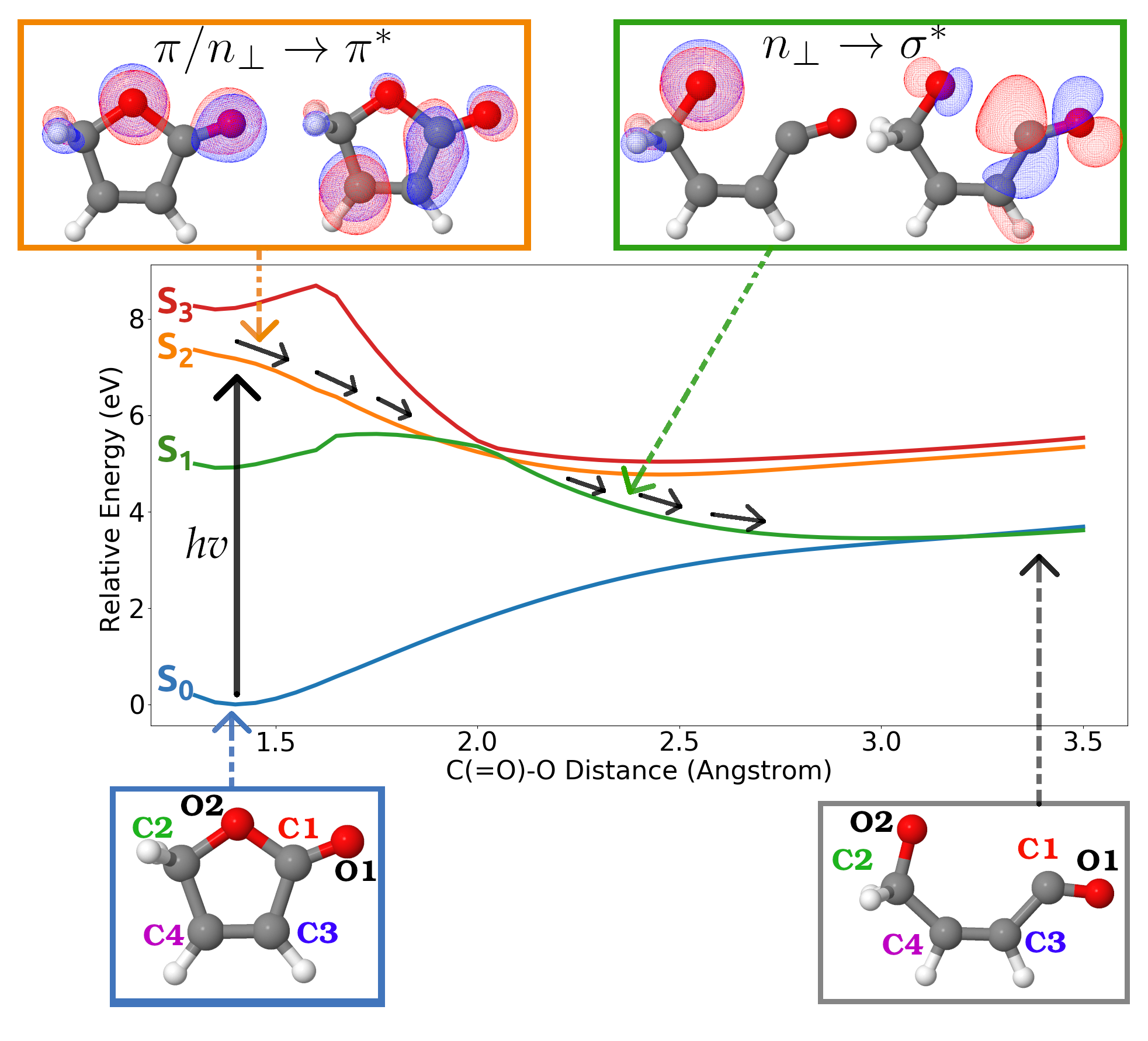}
    \caption{Calculated potential energy slices along the furanone C(=O)-O ring opening coordinate.
    Colored lines are the two lowest adiabatic surfaces of $A'$ (S$_0$ and S$_2$) and $A''$ (S$_1$ and S$_3$) symmetries.
    The anticipated photochemical ring opening process is indicated by solid black arrows (see text for description).  
    The XAS calculations presented here focus only on the S$_1$ surface between 2.35-3.45\AA, as the Lewis structure can be clearly drawn for this surface as a diradical (with unpaired electrons on oxygen 2 and carbon 1).
    The orbitals involved in the primary excitation (i.e. the two SOMOs) on the relevant portions of S$_2$ and S$_1$ surfaces are shown in boxes above the figure, with dashed arrows indicating the plotted geometry.
    The bottom molecules show the ground state equilibrium and ring-open geometries as well as our numbering scheme for unique atoms.
    For all molecular plots, only the sigma bonding network is shown, as the delocalized $\pi$ bonding network changes as a function of ring opening and thus double bonds must be interpreted through $\pi$ orbital occupation numbers. 
    Calculation details and further discussion is included in the SI.
    }
    \label{fig:ring_opening_surfaces}
\end{figure}

A broad range of chemical processes are believed to proceed via short-lived diradical species, including photochemical ring opening\cite{gromov2010theoretical_furanTheoryOpening,murdock2014transient_Furanone_IR_Plus_CASSCF,hua2016monitoring_furanRingOpening_StimulatedXrayRaman,schalk2020competition_Furanone_MD_Plus_TRPES}, reaction intermediates\cite{pedersen1994validity_MolecularBeam_Femtosecond_Diradical} such as those formed in Norrish-Yang\cite{yang1958photochemical_OriginalNorrishYangPaper,chen2016past_NorrishYang_Review} or Patern\'{o}-B\"{u}chi\cite{rykaczewski2020visible_PaternoBuchiRecentWork} mechanisms, and organic photovoltaics \cite{minami2012diradical_PhotovoltaicDiRadicals,niklas2017charge_PhotovoltaicEPR_Review,kanai2010electronic_SomeXrayDiradicalDataOnThinFilms}.
The focus here is on the photochemical ring opening of furanone (see structure in Figure 1), which has been studied theoretically and experimentally in both gas\cite{schalk2020competition_Furanone_MD_Plus_TRPES} and condensed phases\cite{murdock2014transient_Furanone_IR_Plus_CASSCF}.
Calculated potential energy surfaces along the C(=O)-O bond breaking coordinate are shown in Figure \ref{fig:ring_opening_surfaces}, which are in good agreement with previous work\cite{murdock2014transient_Furanone_IR_Plus_CASSCF}.
Briefly, the suspected sub-picosecond ring opening process can be described as an initial photoexcitation to the bright S$_2$ $\pi/n_\perp\rightarrow\pi^*$ surface, followed by a conical intersection onto the S$_1$ $n_\perp\rightarrow\sigma^*$ surface, which severs the C(=O)-O bond.
For a discussion of possible photoproducts, vibrational coupling, and the role of ring puckering, we refer the reader to earlier investigations\cite{murdock2014transient_Furanone_IR_Plus_CASSCF,schalk2020competition_Furanone_MD_Plus_TRPES}.

\begin{figure}[htpb]
    \centering
    \includegraphics[width=\textwidth]{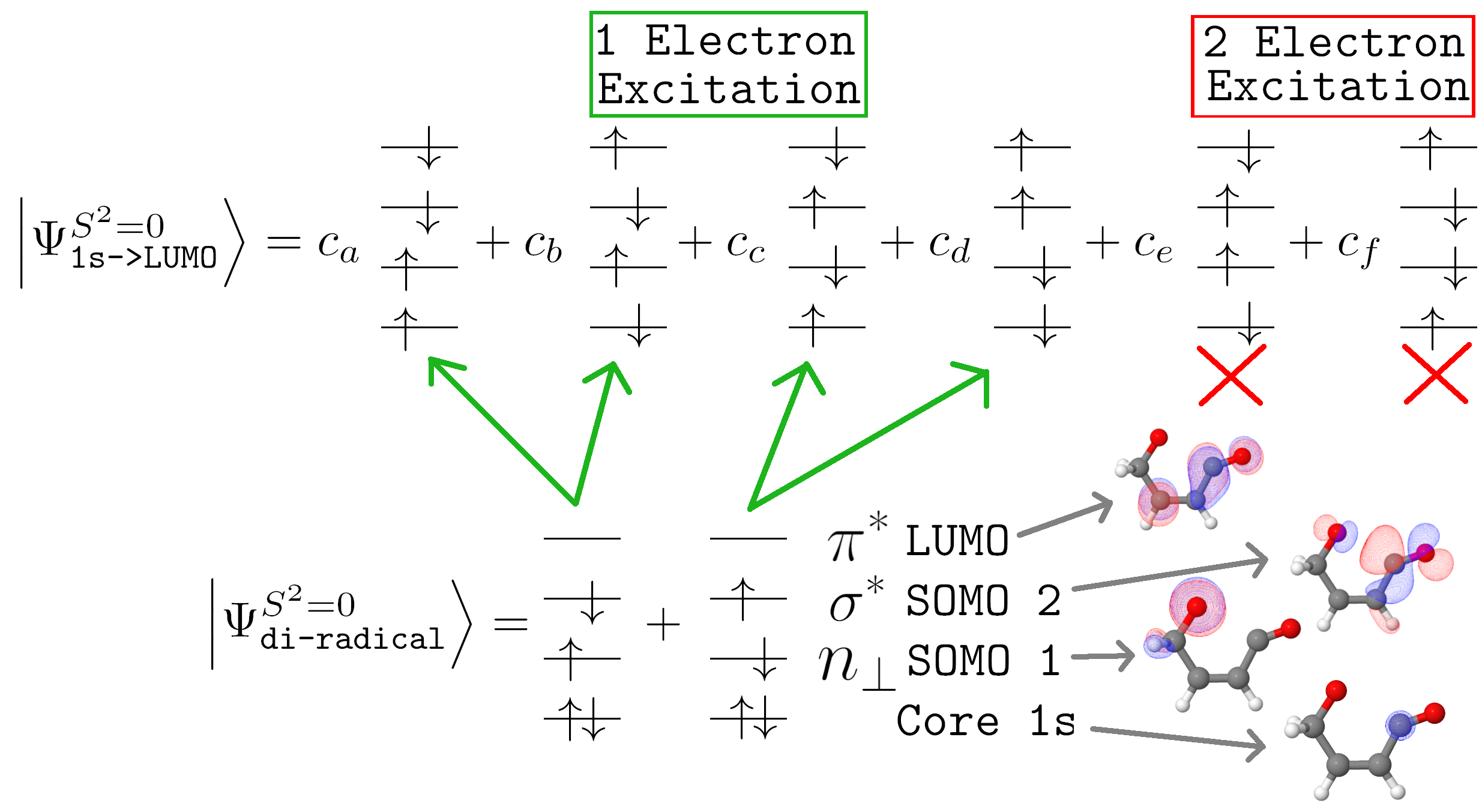}
    \caption{Pictorial MO diagrams for the singlet-coupled diradical state and core-to-LUMO excited state.
    The spin occupancies with coefficients $c_a$ to $c_d$ are all accessible from the diradical via a single electron promotion.
    The spin occupancies with coefficients $c_e$ and $c_f$ are only accessible either via flipping multiple spins or multiple excitations--both of which are disallowed in standard linear X-ray absorption from the C 1s core.}
    \label{fig:keyinsight}
\end{figure}

Observing the ring opening is a prime candidate for probing with time-resolved XAS, as has been done with similar molecules\cite{attar2017femtosecond_XASRingOpening_Cyclohexadiene,bhattacherjee2018photoinduced_FurfuralXASOpening,severino2022non_furanringopenXAS}.
The initial calculations presented here are for the X-ray absorption signature of the ring opening along the S$_1$ $n_\perp\rightarrow\sigma^*$ surface (green in Figure \ref{fig:ring_opening_surfaces}), from 2.35$\text{\AA}$ to 3.45$\text{\AA}$ of the C(=O)-O ring opening coordinate.
Across this geometry range, the diradical SOMOs (plotted in the upper right green box in Figure \ref{fig:ring_opening_surfaces} at $r\text{C(=O)-O}   =2.35$\text{\AA}) are best described as a $n_\perp$ non-bonding p orbital on the in-ring oxygen (oxygen 2) and the $\sigma^*$ orbital that localizes as an in plane p orbital on carbon 1 upon ring opening.
The LUMO is best described as a $\pi^*$ orbital in a delocalized $\pi$ system involving oxygen 1 and carbons 1, 3, and 4 (plotted in Figure \ref{fig:keyinsight}.).
Across this geometry range, the SOMO orbitals go from overlapped with one another to fully spatially separated at the ring open geometry.
While both the carbon and oxygen K-edges are available to be probed, we focus theoretically on the carbon K-edge, as multiple carbons have 1s orbitals with strong spatial overlap with the LUMO $\pi^*$ orbital and therefore are likely to have non-negligible X-ray oscillator strengths for core-to-LUMO transitions.

The lowest energy XAS features, akin to the single radical case, can be expected to be the core-to-SOMO and core-to-LUMO excitations, respectively.
For core-to-SOMO excitations, two unpaired electrons remain in the core-excited probe state -- one resides in the non-re-filled SOMO while the other electron resides in the core orbital excited out of.
These two electrons spin couple into a single  singlet configuration state function (CSF), and from an electronic structure perspective, these core-excited states can be described with single reference theories like ROHF or ROKS\cite{hait2020highly} provided the core hole is enforced with some constraint.
For core-to-LUMO excitations from a singlet diradical, the presence of four unpaired electrons creates the possibility of strongly multi-CSF states.
The singlet spin coupling of these four electrons, or equivalently the coupling of four spin-$\frac{1}{2}$ systems, results in two singlet states that the Hamiltonian can mix to create multi-CSF energy eigenstates.
Under the standard assumptions of linear absorption -- i.e. a single electron is promoted from one orbital to another while conserving spin angular momentum -- spin coupling thus leads to two spectroscopically accessible transitions.
In a minimal picture, each of these singlets can be expressed as a linear combination of the six $\hat{S_z}=0$ ways to singly occupy four orbitals with four electrons, as shown in the top part of Figure \ref{fig:keyinsight}.
A basis of two $\hat{S}^2=0$ CSFs is attainable via Clebsch-Gordon coefficients, however the spectroscopically observable states are the Hamiltonian eigenstates, which are formed from linear combinations of Clebsch-Gordon CSFs.
Obtaining approximate Hamiltonian eigenstates through high-level electronic structure can determine the relative energies of each core-to-SOMO and spin-coupled core-to-LUMO singlet, as well as the approximate X-ray oscillator strength between the core-excited states and diradical valence states.
While core-to-SOMO eigenstates are attainable directly from single reference theories, the multi-CSF nature of the core-to-LUMO eigenstates makes them unattainable with the same methodology without the introduction of further approximations\cite{zhao2021dynamic_DynamicThenStatic}.

\section{Results and Discussion}

To this purpose, we have developed an excited-state-specific electronic structure methodology to capture all core-to-SOMO and core-to-LUMO singlet excited states on equal footing.
Briefly, the central object of our theory is an excited-state-specific orbital optimization for a minimal CSF reference.
We alternate between solving a configuration interaction (CI) problem and relaxing the orbitals for each state specifically using previously developed methodology\cite{tran2019tracking_LANSSCASSCF}.
The minimal reference dictates that each CI stage only diagonalizes the Hamiltonian in the subspace of electron occupations matching the target root of interest.
Using the core-to-LUMO as an example, functionally we diagonalize the 6x6 Hamiltonian formed in the basis of spin occupancies shown in the upper part of Figure \ref{fig:keyinsight}.
All states, including each individual core-to-LUMO spin-coupled component, are calculated separately so each state has its own optimal orbital basis.
Weak correlation and quasi-rediagonalization is achieved via a state--specific selected configuration interaction (sCI)\cite{sharma2017semistochastic_DICE1,holmes2016heat_DICE2,holmes2017excited_DICE_ExcitedStates,smith2017cheap_DICE_Pyscf} treatment.
The sCI states determine the excitation energies and are used to calculate oscillator strengths relative to the core-filled diradical S$_2$ valence state.
The full details of the theoretical method, as well as calculations of the XAS spectra of CO$^+$ to confirm the method's accuracy, are given in the SI.
We note that this methodology shares many similarities with recent work by others focused on the static XAS spectra of the benzyne diradical\cite{Evangalist2023_DSRGXasDiradicals}, for example utilizing a minimal reference orbital relaxation, however the fine details (as dictated in the SI) make a number of different approximations.
Perhaps the most notable differences are state-specific versus state-averaged orbital relaxation and sCI versus perturbative weak correlation treatments.

\begin{figure}[htpb]
    \centering
    \includegraphics[height=0.65\textheight]{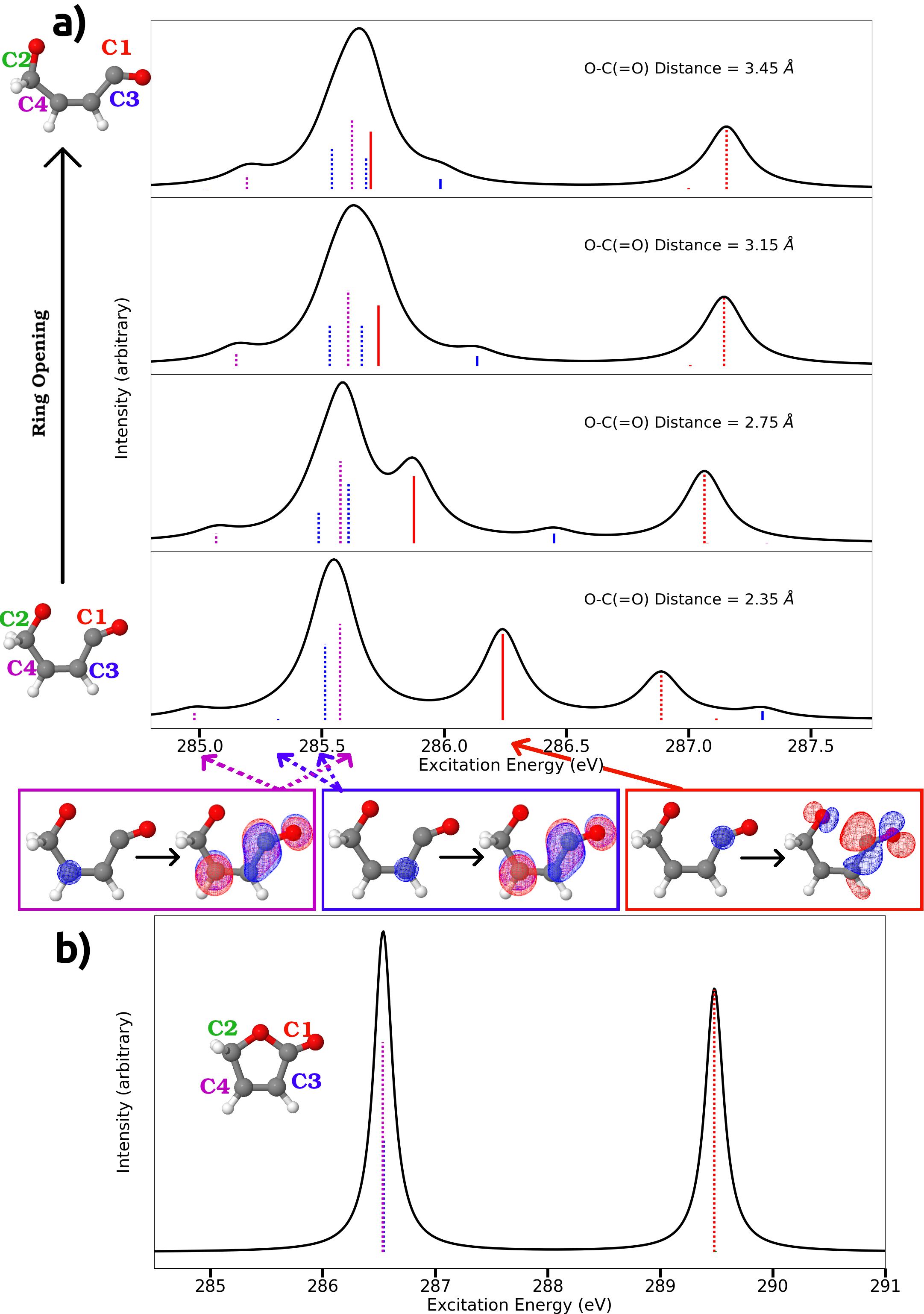}
    \caption{
    \textbf{a)}
    Calculated X-Ray absorption spectra for furanone ring opening at four different geometries.
    Solid lines denote core-to-SOMO transitions in which the final state has two unpaired electrons, while dotted lines denote core-to-LUMO transitions in which the final state has four unpaired electrons.
    The sticks are color coded according to which carbon core orbital is being excited out of.
    The primary states we will discuss are indicated on the bottom spectra with the relevant core (valence) orbitals being excited out of (into).
    Note that while the primary orbital excitations are shown, there are additional unpaired electrons in the final excited states.
    \textbf{b)}
    Calculated XAS for the ground state, ring-closed furanone. 
    Only core-to-LUMO excitations are shown, as excitations to higher lying unoccupied or Rydberg orbitals will likely provide only minor additional structure as in benzene\cite{epshtein2020table_BenzenePlus_Experiment} or furan\cite{severino2022non_furanringopenXAS}.
    Note the excitations from carbons 3 and 4 are nearly degenerate.
    Additional commentary on this spectra is available in the SI.
    The horizontal axes are different for \textbf{a)} and \textbf{b)} so the most important features are clearly visible.
    All states in all spectra are broadened with Lorentzian functions with a width of 0.2 eV, which is a typical spectral resolution of tabletop time resolved XAS setups.
    }
    \label{fig:FullSpectra}
\end{figure}

For each of the four unique furanone carbon atoms, we have calculated four excited states -- namely, core to oxygen 2 $n_\perp$ SOMO1, core to $\sigma^*$ SOMO2, and both of the singlet-coupled core to $\pi^*$ LUMO states (these orbitals are plotted in the lower part of Figure \ref{fig:keyinsight}).
In this study, we focus on the portion of ring opening following internal conversion from the S$_2$ to S$_1$ surface as shown in Figure \ref{fig:ring_opening_surfaces}. 
The XAS spectra probing the dynamics on the initial S$_2$ surface likely involves other dynamic relaxation mechanisms (such as ring puckering\cite{schalk2020competition_Furanone_MD_Plus_TRPES}) and is further complicated by the delocalized $\pi$ nature of the unpaired electrons involved in the primary excitation.
We therefore choose to only focus on the S$_1$ surface at distances greater than 2.35\AA$\text{}$ as after this point the diradical electrons reside in spatially localized orbitals, which allows for a clearer analysis of spin coupling effects on the XAS spectra.
These sixteen states are calculated at four different geometries, and the calculated X-ray spectra are presented in Figure \ref{fig:FullSpectra} a).
Additionally, the core-to-LUMO excitation for each of the four unique carbon atoms at the equilibrium ground state geometry are calculated and the resulting spectra is shown in Figure \ref{fig:FullSpectra} b), for reference.
Due to the ``local selection rule" nature of core excitations\cite{manne1970molecular_LocalSelectionRules_XAS}, many of these states have negligible intensity.
Intuitively, this is because the core orbital has negligible spatial overlap with any of the SOMO or LUMO orbitals.
The carbon 2 core 1s orbital, for example, does not spatially overlap with either of the SOMO orbitals or the LUMO orbital, and therefore while all core-to-SOMO1, core-to-SOMO2, and both core-to-LUMO spin-coupled excitations for carbon 2 are calculated and plotted, none of these transitions have appreciable X-ray oscillator strengths and thus are not visible in Figure \ref{fig:FullSpectra} a).
Excitations into the oxygen 2 $n_\perp$ SOMO1 may be expected to have the lowest overall excitation energies, as this SOMO is doubly occupied in the Aufbau configuration.
Indeed, for carbons 2, 3, and 4, the calculated excitation energies for all of these transitions are $<$285.1 eV.
However, as none of the carbon core orbitals spatially overlap with this SOMO, these excitations all have negligible intensity.

The most intense feature that changes as a function of geometry is the red-shift of the carbon 1 core to $\sigma^*$ SOMO2 excitation (solid red stick spectra, Figure \ref{fig:FullSpectra} a)) from about 286.25 to 285.75 eV.
The red-shift can be understood as a consequence of the bond breaking.
As the ring opens, the SOMO that is being probed transforms from having primarily $\sigma^*$ anti-bonding character between carbon 1 and oxygen 2 to having 2p like non-bonding character localized on carbon 1.
This change in orbital character creates a new non-bonding vacancy to excite the carbon 1 atom's core electron into, red-shifting the peak as the ring opens.
The only other carbon core to $\sigma^*$ SOMO2 excitation with noticeable intensity is the carbon 3 core to $\sigma^*$ SOMO2 excitation (solid blue stick spectra, Figure \ref{fig:FullSpectra} a)), which red-shifts from about 287.3 eV to 286.0 eV.
This feature is however much weaker than the carbon 1 core to $\sigma^*$ SOMO2 transition.
Additionally, the magnitude of our reported red-shifts are again only for the portions of the ring opening following internal conversion from the S$_2$ to S$_1$ surface.
The complete time-resolved experiment may identify a much larger red-shift of the core-to-$\sigma^*$ excitation energies as the $\sigma^*$ orbital is very high energy at the short C(=O)-O distances involved on the S$_2$ surface.
Excitations from the carbon 2 and 4 core orbitals to the $\sigma^*$ SOMO2 do not have appreciable oscillator strengths.
As such, all other features shown in Figure \ref{fig:FullSpectra} a) are attributable to core-to-LUMO excitations.
Importantly, this means utilizing any secondary features in the spectra that could uniquely identify this ring opening coordinate, as opposed to other relaxation pathways, requires analyzing the two resulting states formed from spin-coupling in the core-to-LUMO excited states.

The dotted red line, which blue-shifts from about 286.9 to 287.2 eV, is one of the two components of the carbon 1 core-to-LUMO excitation.
Both spin-coupled components are calculated and plotted, however only one of the components has appreciable oscillator strength at any geometry studied.
As we have only calculated the excitations to the SOMO and LUMO orbitals, it is possible this state would be obfuscated by excitations to higher lying virtual or Rydberg transitions from other carbon atoms.
As such, we turn our attention in detail to the feature at about 285.5 eV, which remains relatively constant as a function of geometry but has an increasing and blue-shifting low energy shoulder at 285 eV.
These features are composed of excitations from the backbone carbons 3 and 4 cores to the $\pi^*$ LUMO, and therefore involve two spin-coupled components per carbon.
While the increasing shoulder is relatively weak, our calculations show it is the lowest energy observable transition in the total spectra and therefore, without other overlapping transitions, may be viable to measure experimentally.

\begin{figure}[htpb]
    \centering
    \includegraphics[width=\textwidth]{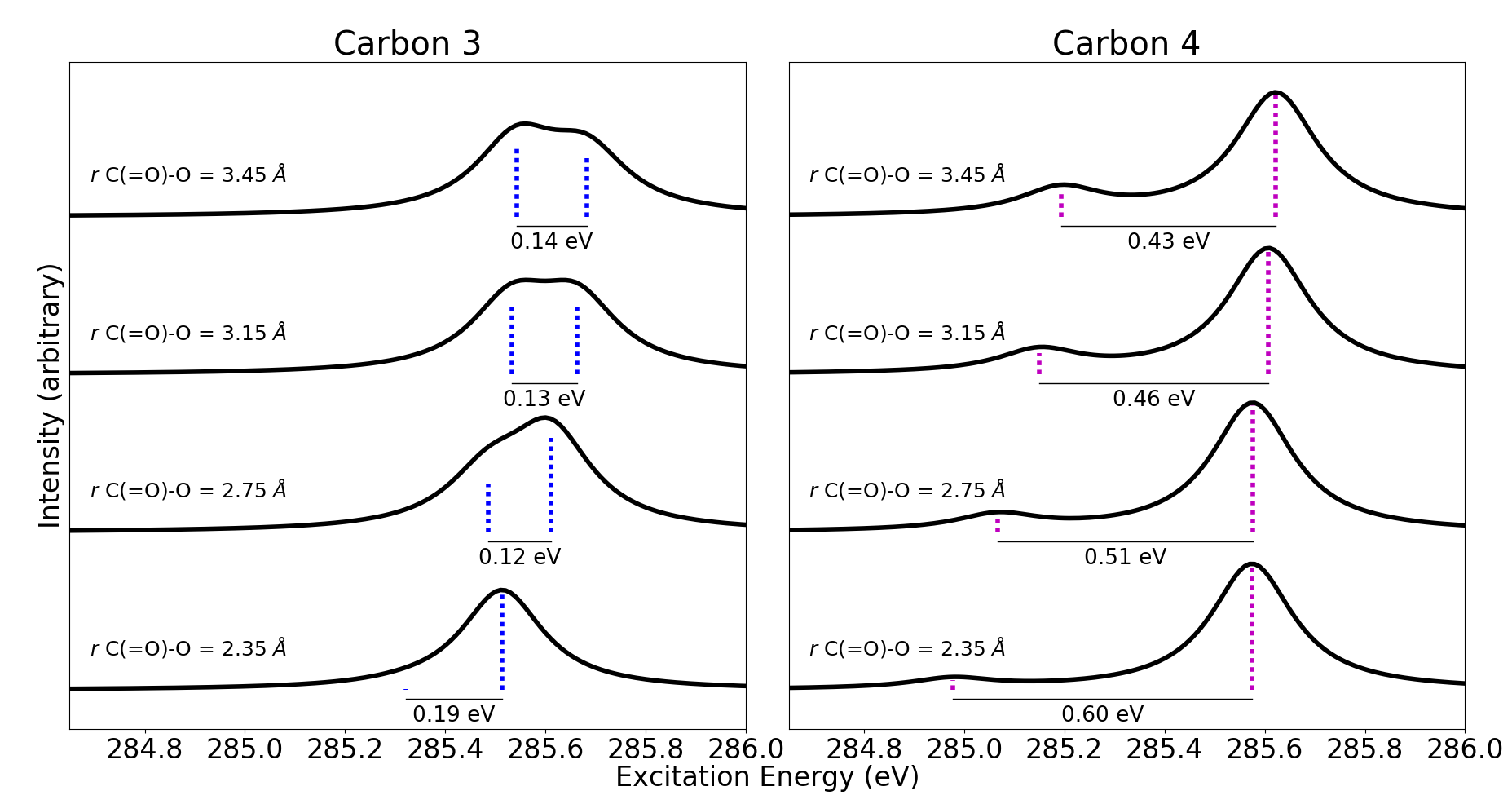}
    \caption{
    Core-to-LUMO spin-coupled transitions for carbons 3 and 4 as a function of ring opening geometry.
    The dotted blue and magenta sticks are identical those that are plotted in Figure \ref{fig:FullSpectra} a), but isolated here for clarity in analyzing the core-to-LUMO spin-coupled probe states.
    The spin splitting is indicated for each pair of spin-coupled core-to-LUMO transitions.
    All sticks are broadened with Lorentzian functions with width of 0.2 eV.
    }
    \label{fig:C3C4LUMOISO}
\end{figure}

To further investigate the effect of the spin coupling on the core-to-LUMO transitions, the two spin-coupled core-to-LUMO transitions for carbons 3 and 4, which create this low energy feature and blue-shifting shoulder, have been isolated and clearly plotted for each carbon in Figure \ref{fig:C3C4LUMOISO}.
Starting with carbon 3, for all geometries considered, the spin splitting remains small, $<$0.2 eV.
While the absolute oscillator strengths of the two components are not constant, when convoluting these states with Lorentzian functions with FWHM=0.2eV (consistent with previous studies\cite{vidal2019new_fcCVSEOMCCSD} and typical of laboratory table top spectral resolutions), no clear spin-splitting features or shifts will be easily observed at any geometry.
That is, in the calculated spectra for the two spin-coupled peaks described by the carbon 3 core to $\pi^*$ LUMO, we predict minimal discernible geometry effects and therefore few changes in the time-dependent XAS absorption signal, likely precluding these transitions from being utilized to uniquely identify the ring opening coordinate.
A detailed discussion of the individual component's absolute intensities is presented in the SI.

For carbon 4, however, the spin splitting is much larger as the ring is beginning to break open -- about 0.60 eV at the C(=O)-O distance of 2.35\AA.
As the bond breaks, the spin splitting decreases to 0.43 eV.
As a function of geometry, the excitation energy to the higher energy spin-coupled component remains about constant, while the excitation energy to the lower energy component blue-shifts towards the higher state.
Furthermore, the calculated oscillator strength of the higher energy component remains about constant, while the lower energy component's calculated oscillator strength noticeably increases as this component blue-shifts by about 0.2 eV over the course of the ring opening.
Interestingly, this is about the same magnitude blue-shift as calculated for the one component of the carbon 1 core-to-LUMO excitation (dotted red line, Figure \ref{fig:FullSpectra} a)) over the geometries studied on the S$_1$ surface, which may imply this is the approximate energy for destabilizing the LUMO over this geometry coordinate.
However, carbon 1 is also breaking a bond to oxygen 1, which should affect the carbon atom's core orbital energy, and these states are all influenced by the spin coupling of the four unpaired electrons, making direct commentary on the LUMO difficult.
Nevertheless, the growth in oscillator strength and blue-shift of this lower energy carbon 4 core-to-LUMO spin-coupled excitation is clearly dependent on geometry, and therefore will yield a time-dependent XAS absorption signal, 
Note that all the features at 285 eV due to carbon 4 are weaker than several main features (Figure \ref{fig:FullSpectra} a)), but the subtle geometrical information contained in the secondary carbon sites during ring opening, and the spin splittings involving four unpaired electrons, both embody new aspects of transient X-ray spectroscopy of diradicals that have not been previously explored.

To summarize, over the portions of the ring opening coordinate studied, our calculations predict no discernible (within typical experimental broadening) spin split or geometry dependent signal for carbon 3 core-to-LUMO excitations, while carbon 4 shows greater spin splitting with the lower energy component gaining intensity and blue-shifting as the ring opens.
This is surprising, in that both of these carbons can be described as backbone carbons, neither of which is breaking or forming bonds as the ring opens, and yet they display qualitatively important geometry dependent XAS features.

The qualitative difference between carbon 3 and carbon 4 is a consequence of the details of spin coupling between the four unpaired electrons in the highly localized core, oxygen 2 $n_\perp$ SOMO1, $\sigma^*$ SOMO2, and $\pi^*$ LUMO orbitals.
When the bond breaks, the SOMOs localize to either end of the ring open photoproduct, resulting in a diradical in which the two electrons reside on separate atoms.
As we will see, this spatial separation modifies the details of the spin coupling in ways that lead to qualitatively different XAS core-to-LUMO signatures for these two carbons.

To investigate how these changes in spin coupling impact the XAS signals, we return our focus to the unpaired electrons in the diradical and core-excited states.
Inspecting the six $\hat{S_z}=0$ spin occupancies that compose the core-to-LUMO transitions (upper part of Figure \ref{fig:keyinsight}), there is a key observation.
Due to the structure of the initial singlet diradical state and the fact that the optically allowed excitation process preserves the $S_z$ spin of the transitioning electron, the electrons in the SOMO1 and SOMO2 orbitals must necessarily have opposite spins in order for a configuration to make a nonzero contribution to the oscillator strength.
In the first four core-excited state spin occupancies (coefficients $c_a$ to $c_d$) the two SOMO electrons have opposite spins.
Again relying on the standard assumptions of linear absorption, each of these occupancies is an allowed transition via promotion of a single core electron to the unoccupied orbital.
Importantly, in the final two spin occupancies (with coefficients $c_e$ and $c_f$), the two SOMO electrons have the same spin, meaning, relative to either occupancy of the singlet diradical, reaching these occupancies requires two electron promotions.
These observations imply that the first four spin occupancies are reachable via a dipole allowed excitation in the diradical whereas excitations to the final two spin occupancies are dipole forbidden.

Based on these observations, we can conjecture a simple spin-occupancy-induced selection rule: a singlet composed of primarily dipole allowed spin occupancies is likely spectroscopically bright, whereas a singlet composed of mainly dipole forbidden spin occupancies is likely spectroscopically dark. 
This spin-occupancy selection rule is fundamentally built upon two approximations.
First, building upon Fermi's Golden Rule, we only consider linear absorption, where a single photon excites a single core electron into an unoccupied orbital.
Second, the photon only acts on the electronic degrees of freedom and does not interact with the spin degrees of freedom.
These approximations are equivalent to neglecting higher than linear order absorption processes and all relativistic light-matter interactions, respectively.
The prime candidates for experimentally adhering to this spin-occupancy selection rule are thus second period elements due to small relativistic effects\cite{takahashi2017relativistic_RelativisticEffects} with observability on K-edge transitions with ultra-fast tabletop spectrometers.

\begin{figure}[htpb]
    \centering
    \includegraphics[width=\textwidth]{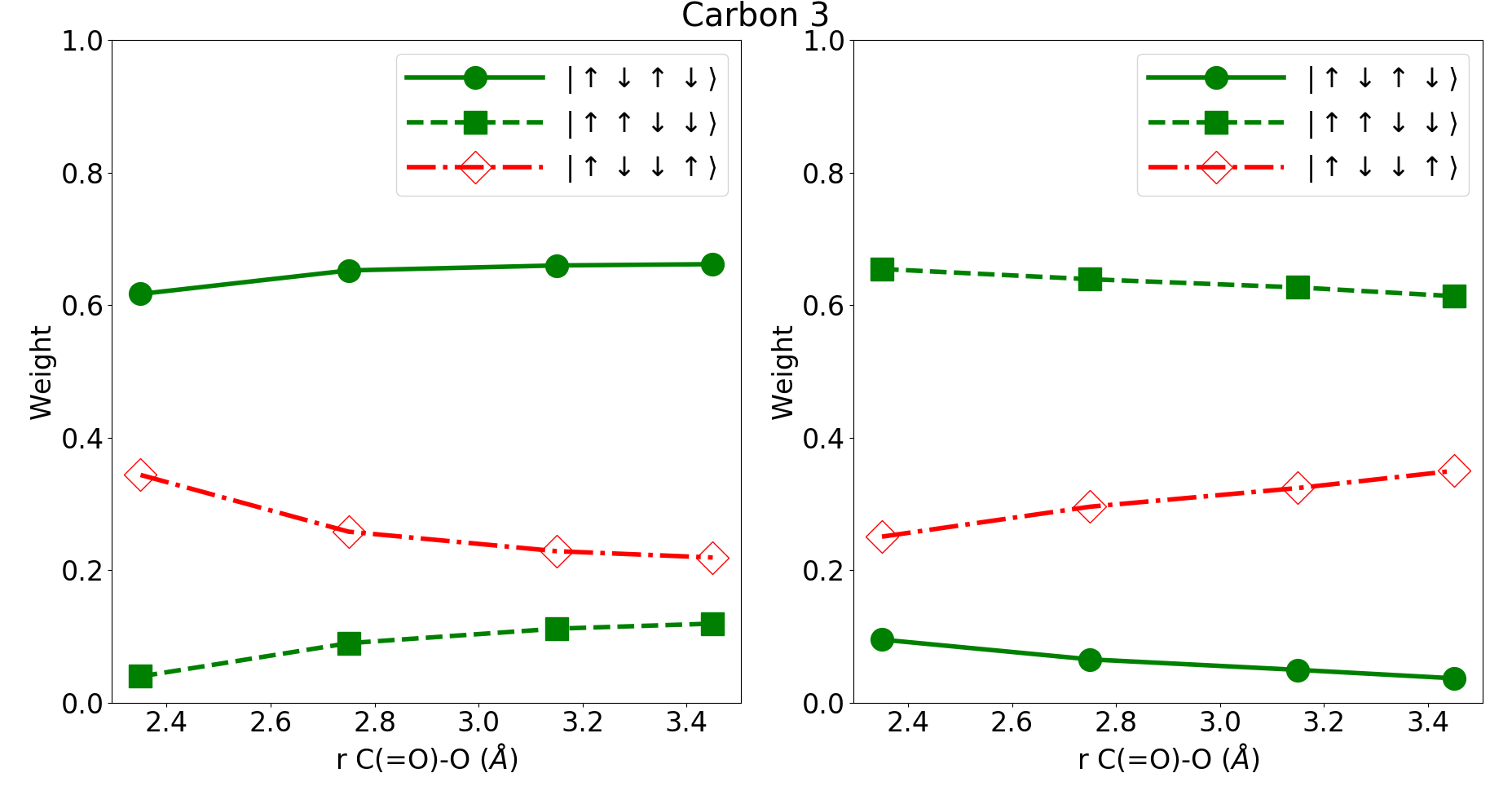}
    \caption{
    Contributions of X-ray optically allowed (green) and disallowed (red) spin occupancies to the minimal CSF carbon 3 core-to-LUMO Hamiltonian eigenstates as a function of ring opening.
    The left (right) plot shows the composition of the lower (higher) energy eigenstate.
    The full theoretical predictions for these transitions are shown in Figure \ref{fig:C3C4LUMOISO}, left.
    The plotted quantity is twice the sum of the coefficient squared for the listed determinant.
    For example, the solid line (notated \abab) can be interpreted as 2$|c_c|^2$ (with $c_c$ taken from Figure \ref{fig:keyinsight}), or equivalently as $|c_b|^2+|c_c|^2$ as these spin occupancies necessarily have the same magnitude for a singlet.
    }
    \label{fig:C3StateComp}
\end{figure}

\begin{figure}[htpb]
    \centering
    \includegraphics[width=\textwidth]{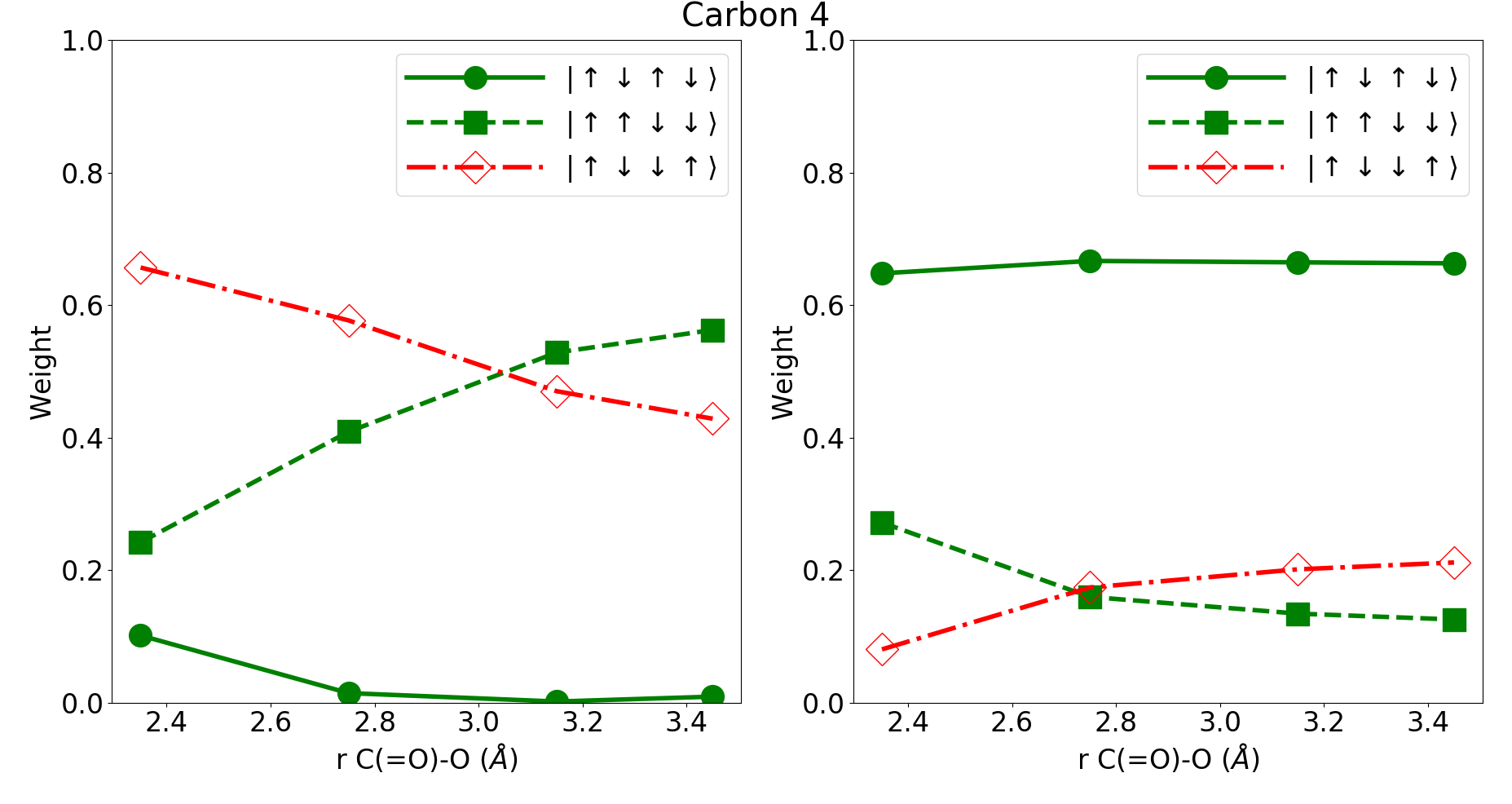}
    \caption{
    Contributions of optically allowed (green) and disallowed (red) spin occupancies to the minimal CSF carbon 4 core-to-LUMO Hamiltonian eigenstates as a function of ring opening.
    The left (right) plot shows the composition of the lower (higher) energy eigenstate.
    The full theoretical predictions for these transitions are shown in Figure \ref{fig:C3C4LUMOISO}, right.
    The interpretation is the same as in Figure \ref{fig:C3StateComp}.
    }
    \label{fig:C4StateComp}
\end{figure}

To check the validity of this rule, we return to our calculated carbon 3 and carbon 4 core-to-LUMO excited states for furanone.
We plot the individual spin occupancy compositions for both singlet spin-coupled core-to-LUMO states for each carbon as a function of the ring opening coordinate in Figures \ref{fig:C3StateComp} and \ref{fig:C4StateComp}.
It is important to note that these compositions are taken from the minimal CSF orbital optimization stage, and not from our final sCI wavefunctions.
While the final sCI wavefunctions are what determine excitation energies and oscillator strengths, this intermediate orbital optimized minimal CSF wavefunction is a simpler entity to interpret, more easily attained from a practical perspective, and can be analyzed in a way that is similar to the approach taken by others\cite{hait2020accurate_Dip_ROKSCoreRadicalRemixing,zhao2021dynamic_DynamicThenStatic} for doublet state single radical species.

Starting with carbon 3 (Figure \ref{fig:C3StateComp}), for each of the spin-coupled components, no matter the geometry, both states are dominated by X-ray optically allowed spin occupancies.
This is qualitatively consistent with the peaks calculated from the weakly correlated sCI wavefunction as shown in Figure \ref{fig:C3C4LUMOISO}.
Across the ring opening coordinate, there is no predicted qualitative change in the total signal (again, see the SI for a more detailed discussion of absolute intensities).
For carbon 4 (Figure \ref{fig:C4StateComp}), the higher energy spin-coupled component is dominated by X-ray optically allowed spin occupancies at all geometries.
The lower energy component, however, is dominated by the optically disallowed spin occupancies at short C(=O)-O distances but becomes composed of roughly equal weights of optically allowed and disallowed spin occupancies at large C(=O)-O distances.
Again, this is in qualitative agreement with the calculated spectra as plotted in Figure \ref{fig:C3C4LUMOISO}.
The higher energy component has significant oscillator strength independent of geometry, while the lower energy component is overall weaker but gains oscillator strength as the ring opens.

This simple metric of investigating the composition of optically allowed versus disallowed spin occupancies provides the framework we will use to understand how a change in geometry can lead to a qualitative change in an individual carbon atom's core-to-LUMO XAS signatures due to spin coupling.
Quantitative predictions of the magnitude of spin splitting and absolute intensities still require high-level calculation.
However, as the data suggest, a qualitative understanding of the geometry dependence of XAS core-to-LUMO features is possible if one has an idea of the minimal reference state.
Focusing on the spin coupling of the four unpaired electrons, the minimal reference states can be determined by solving a geometry dependent, Heisenberg spin Hamiltonian\cite{ciofini2005mapping_PostHF_SpinSolving}
\begin{equation}
    \hat{H}=-\sum_{i,j}^4J_{ij}(\vec{R}) \vec{S_i} \cdot \vec{S_j}
\end{equation}
where $J_{ij}(\vec{R})$ is the geometry dependent exchange coupling between orbitals $i$ and $j$.
Given a set of exchange couplings, $J_{ij}(\vec{R})$, the singlet eigenstates are easily attained by diagonalization of a 6x6 matrix--the explicit form of which is available in the SI.

Estimates for $J_{ij}(\vec{R})$ for core-excited states are calculable via sophisticated electronic structure\cite{hait2020accurate_Dip_ROKSCoreRadicalRemixing,zhao2021dynamic_DynamicThenStatic,Evangalist2023_DSRGXasDiradicals}.
However, we find that utilizing a H\"{u}ckel-like approach to estimate these couplings is sufficient to aid in understanding the overall geometry dependence.
Similar to H\"{u}ckel theory, we make the simplifying assumption that $J_{ij}$ elements are ``on" or ``off" depending on the spatial overlap of the two orbitals involved.
We make this approximation on the basis that exchange couplings decay exponentially with distance\cite{schwegler1996linear_ExponentialDecayOfExchange} -- indeed an example of this exponential decay is shown in the SI.
We say that an exchange coupling is ``on" (ie, $J_{ij}\neq 0$) if two orbitals overlap on the same atom or if they have a nearest neighbor interaction, otherwise we say the coupling is ``off" (ie $J_{ij}=0$).
While these approximations are qualitative, we are primarily interested in the relative pictures of X-ray optical activity that these estimates provide across a geometry coordinate, and therefore it is not the exact values of couplings that are of utmost importance, but rather how two sets of couplings -- $J_{ij}(\vec{R})$ and $ J_{ij}(\vec{R'})$ -- differ from one another over the course of a chemical reaction.
By approximating exchange couplings at multiple geometries via these H\"{u}ckel-like rules, it is straightforward to obtain qualitative estimates of how different core-to-LUMO transition intensities are affected by changes in geometry by investigating the spin Hamiltonian's singlet eigenstates spin occupancies.

\begin{figure}
    \centering
    \begin{subfigure}{0.44\textwidth}
        \centering
        \includegraphics[width=\textwidth]{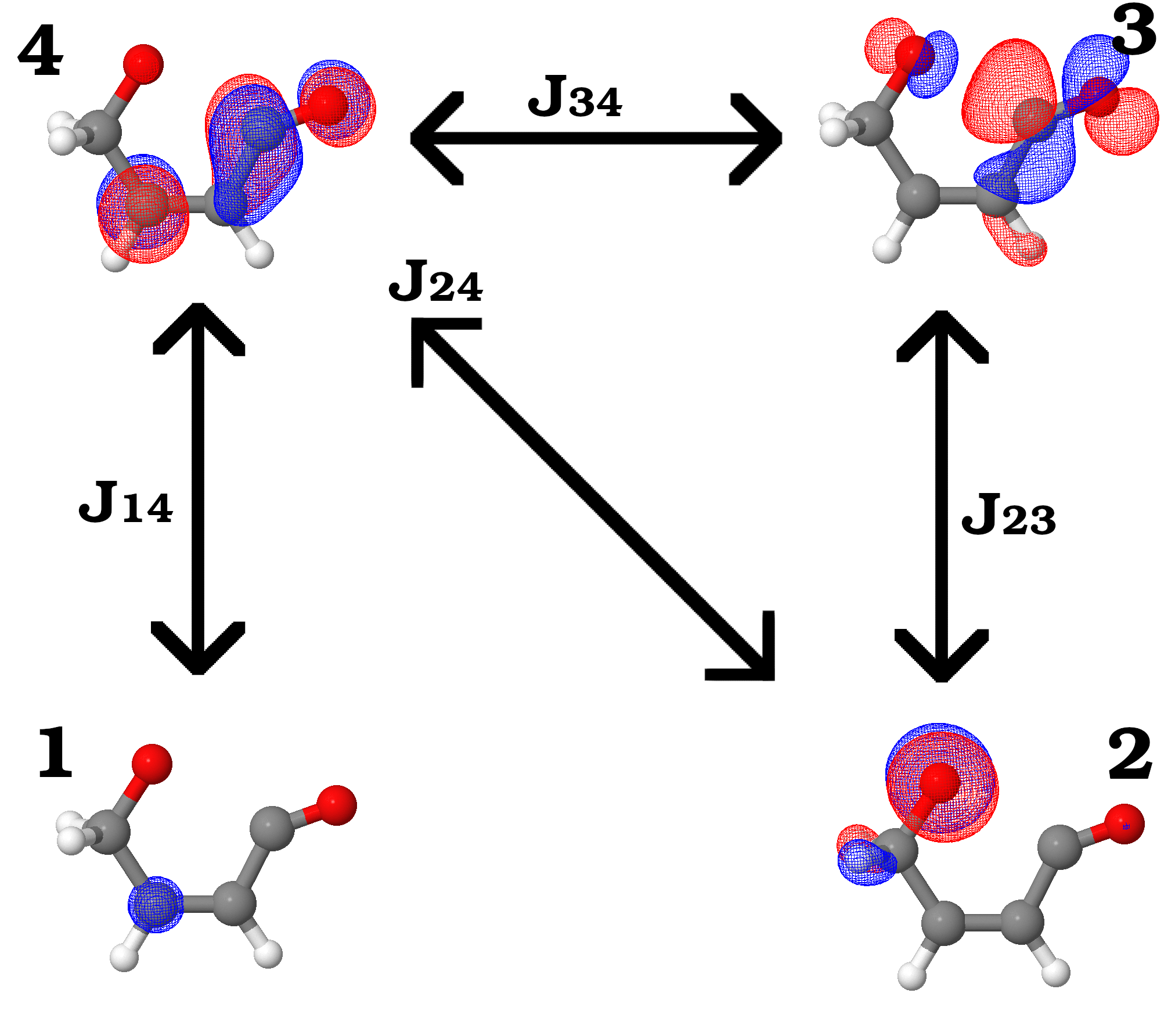} 
        \caption{$r\text{C(=O)-O} = 2.35$\AA}
    \end{subfigure}\hfill
    \begin{subfigure}{0.44\textwidth}
        \centering
        \includegraphics[width=\textwidth]{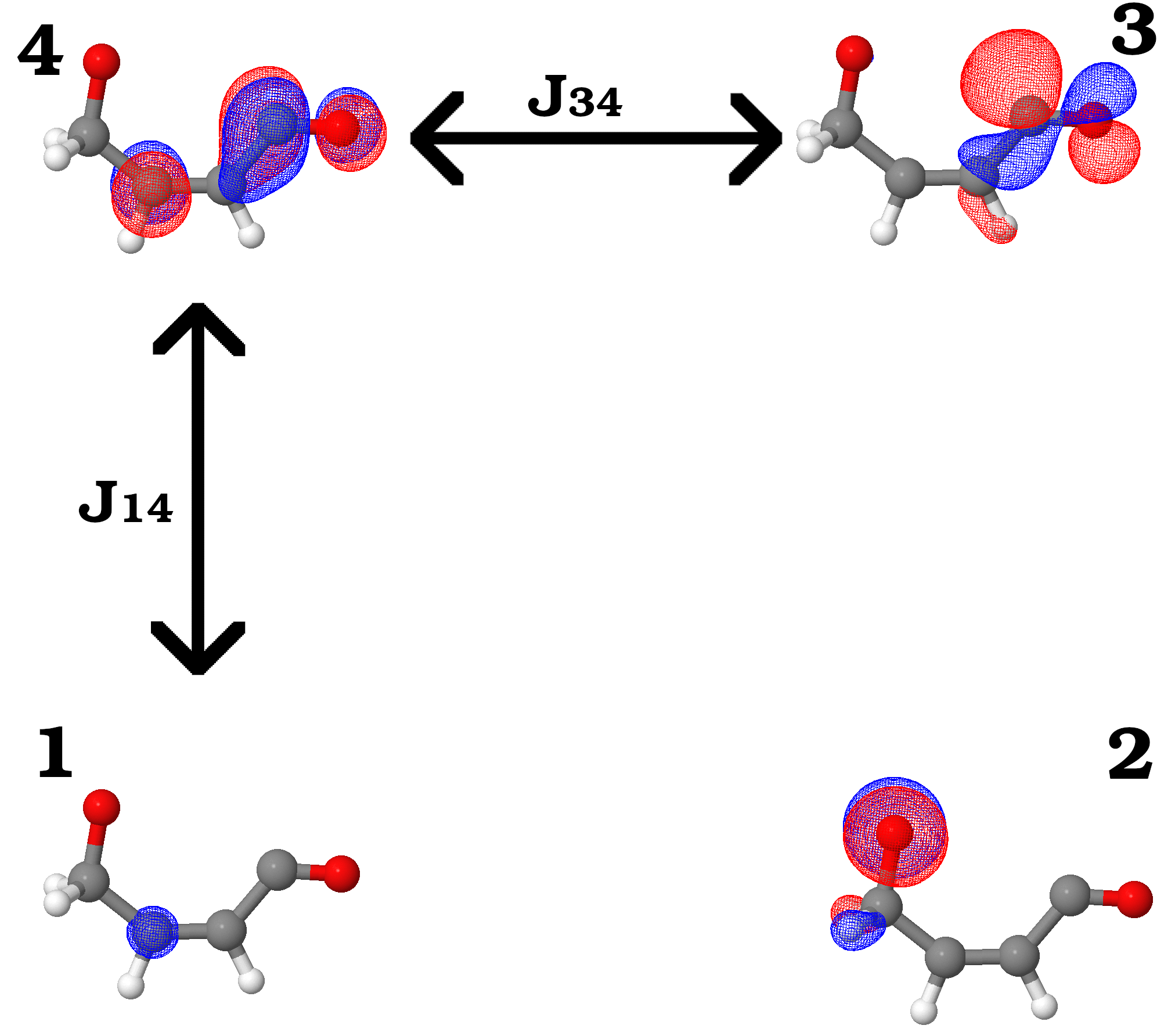} 
        \caption{$r\text{C(=O)-O} = 3.45$\AA}    
    \end{subfigure}
    \caption{
    Exchange couplings for carbon 4 in the furanone photochemical ring opening.  
    The orbitals are labeled as follows: core 1s is orbital 1, the oxygen $n_\perp$ SOMO1 is orbital 2, the $\sigma^*$ SOMO2 is orbital 3, and the $\pi^*$ LUMO is orbital 4.
    Couplings that are considered ``on" at a given geometry are shown with arrows indicating which two orbitals the coupling is between.}
    \label{fig:C4SquareExchange}
\end{figure}

Returning to the furanone ring opening, we start by applying this analysis to carbon 4, as this carbon has the interesting geometry dependent shoulder shown in Figure \ref{fig:C3C4LUMOISO}.
The relevant exchange couplings to consider are shown in Figure \ref{fig:C4SquareExchange}.
When the ring is just starting to open (Figure \ref{fig:C4SquareExchange} (a)), $J_{14}$ and $J_{34}$ are considered strong couplings as they have direct spatial overlap.
$J_{24}$ and $J_{34}$ are less clear, but we rationalize that as the bond hasn't fully severed yet, the interaction is likely still strong through the space across the breaking bond.
The other two couplings $J_{12}$ and $J_{13}$ are considered off since the orbitals involved are spatially separated by more than one atom.
Once the bond has severed (Figure \ref{fig:C4SquareExchange} (b)), $J_{24}$ and $J_{34}$ are now considered off, as the oxygen $n_\perp$ is now spatially distanced from the $\sigma^*$ ($J_{23}$) and $\pi^*$ ($J_{24}$) orbitals.
These sets of exchange couplings and resulting singlet eigenstate spin occupancy compositions from diagonalizing the spin Hamiltonians are presented in Table \ref{tab:c4_huckel}.

\begin{table}[hptb]
    \centering
    \begin{tabular}{l |c|c}
         \hline
         \multicolumn{3}{c}{$r\text{C(=O)-O}=2.35\text{\AA}$} \\
         \hline 
         \multicolumn{3}{c}{$J_{14}=J_{23}=J_{24}=J_{34}\neq 0$,$\quad$ $J_{12}=J_{13}=0$} \\
         \hline
         Component & $c_{bright}$ & $c_{dark}$ \\
         \hline
         Higher Energy & 1.00 & 0.00 \\
         Lower Energy & 0.33 & 0.67 \\
         \hline
         \hline
         \multicolumn{3}{c}{$r\text{C(=O)-O}=3.45\text{\AA}$} \\
         \hline
         \multicolumn{3}{c}{$J_{14}=J_{34}\neq0$, $\quad$$J_{12}=J_{13}=J_{23}=J_{24} = 0$}\\
         \hline
         Component & $c_{bright}$ & $c_{dark}$ \\
         \hline
         Higher Energy & 0.83 & 0.17 \\
         Lower Energy & 0.50 & 0.50 \\
         \hline
    \end{tabular}
    \caption{
    Relative weights of optically allowed (defined as $c_{bright}\equiv |c_a|^2+|c_b|^2+|c_c|^2+|c_d|^2$, with $c_{a-d}$ defined in Figure \ref{fig:keyinsight}) versus disallowed (defined as $c_{dark}\equiv|c_e|^2+|c_f|^2$) spin occupancies of the spin Hamiltonian singlet eigenstates for carbon 4 core-to-LUMO transitions formed by estimating couplings with H\"{u}ckel-like approximations.}
    \label{tab:c4_huckel}
\end{table}

At $r \text{C(=O)-O}=2.35$\AA, the expected XAS signal for the core-to-LUMO excitation would contain a single feature; namely the higher energy spin-coupled component since it is dominated by X-ray optically allowed spin occupancies (Upper Energy $c_{bright}=1.0$).
The lower energy component is expected to be X-ray optically dark (or at least present as only a minor feature) since it is dominated by optically disallowed occupancies (Lower Energy $c_{dark}=0.67$).
Once the ring opens, the higher energy component remains dominated by optically allowed spin occupancies (Upper Energy $c_{bright}=0.83$), while the lower energy singlet gains intensity, being composed of equal parts optically allowed and disallowed spin occupancies (Lower Energy $c_{bright}=c_{dark}=0.5$).
Remarkably, this is the interpretation that is arrived at by electronic structure when using either the fully detailed sCI (Figure \ref{fig:FullSpectra} b)) or the minimal reference (Figure \ref{fig:C4StateComp}) wavefunctions.

For carbon 3, which does not show a geometry dependent signal, a similar H\"{u}ckel-like analysis is possible.
The relevant couplings and results of diagonalizing the spin Hamiltonians are provided in the SI.
The couplings are similar to carbon 4, with the primary difference being the coupling between the core and $\sigma^*$ SOMO ($J_{13}$) is a nearest neighbor interaction and considered ``on" at all geometries.
At the H\"{u}ckel-like level of approximation, the most useful result is that at the ring open geometry the singlet eigenstates are degenerate, and therefore this model predicts potentially minimal spin splitting.
The spin splitting observed in the complete calculations (Figure \ref{fig:C3C4LUMOISO}) is therefore a consequence of the fine details of the spin coupling -- that is to say we are reminded that the H\"{u}ckel-like approximations are too simple for quantitative energetics.
Further, no clear predictions can be made about a geometry dependent signal from the spin eigenstates.
Importantly however, from the H\"{u}ckel perspective, we can understand that because of the single additional strong coupling ($J_{13}$), there may be a qualitative difference in the XAS signatures of carbons 3 and 4.
Discussion of applying this H\"{u}ckel-like model to carbons 1 and 2 is presented in the SI.

\section{Conclusions}

To summarize, through a spin occupancy induced selection rule and diagonalizing a spin Hamiltonian formed with H\"{u}ckel-like approximations, we can qualitatively understand a geometry dependent signal arrived at by high-level calculation.
The theoretical results suggest these approximations may be most applicable when the spin splitting is large (as is the case for carbon 4), whereas when the splitting is small, detailed electronic structure is required to resolve the spin coupling details (as is the case for carbon 3).
For the furanone ring opening studied here, these H\"{u}ckel-like approximations reveal that the growth of a carbon 4 core-to-LUMO spin-coupled component is a result of spatially distancing the oxygen $n_\perp$ SOMO from the $\sigma^*$ SOMO and $\pi^*$ LUMO in the ring open product.
Importantly, in addition the core-to-SOMO features that can provide local insights to chemical dynamics, this core-to-LUMO transition can serve as a secondary, non-local reporter of the bond breaking -- in this case induced by spin coupling through a delocalized $\pi^*$ system.

Moving forward, it will be interesting to study different atomic edges where the magnitude of spin splitting may be different (for example, the oxygen K-edge spin splitting is much larger than the carbon K-edge in the carbon monoxide radical\cite{couto2020carbon_COPlus_XAS_Spectra}).
Further, it will be interesting to investigate whether the spin-occupancy-induced selection rule introduced here can lead to more general predictions about the structure of spin-coupled core-to-unoccupied excitations -- for example if the spin splitting is large it may be possible to typically anticipate a doublet structure with one intense and one weak component. 
Interestingly, in work by others on the static XAS spectra of the benzyne\cite{Evangalist2023_DSRGXasDiradicals} diradical, when including core-to-unoccupied transitions, the reported spectra appear to have many transitions with this general doublet structure; however, as they mention, these states are not as straightforward to analyze due to their potentially multi-electron nature.
The results presented here suggest chemical dynamics that involve the spatial separation of the two diradical electrons may be an interesting avenue to pursue to observe these signals in time-resolved experiments, as the fundamentally local effects of electron exchange will decay as electrons are separated and dynamic spin-coupled XAS features may result.
However, to the best of our knowledge, no XAS experiment has yet clearly resolved these features, and new experiments will be important to test the theoretical predictions.

\pagebreak

\begin{suppinfo}

\setcounter{figure}{0}
\renewcommand{\figurename}{Fig.}
\renewcommand{\thefigure}{S\arabic{figure}}

\setcounter{table}{0}
\renewcommand{\tablename}{Table}
\renewcommand{\thetable}{S\arabic{table}}
\renewcommand{\thepage}{S\arabic{page}}

\section{State-averaged CASSCF PES scans}
The potential energy surface scans along the C(=O)-O ring opening coordinate are calculated via state-averaged CASSCF.
Starting from the geometry given in Table \ref{tab:SI:middlegeom} (at C(=O)-O distance 1.75\AA), forward and backward surface scans are performed between C(=O)-O distances of 1.3 to 3.5\AA $ $ with steps of 0.05 \AA. 
At each step, the C(=O)-O distance is fixed and the remaining internal coordinates are optimized according the lowest energy totally symmetric state's gradients calculated with a 4-state state-averaged 10 electron, 8 orbital (10e,8o) CASSCF calculation in the 6-31+G* basis.
The active space at the 1.75\angs geometry is shown in Figure \ref{fig:SI:SACASORBS}.
$C_s$ symmetry is used, and the 4-states are calculated as the lowest two energy states in each irreducible representation.
Calculations are performed using MOLPRO\cite{werner2012molpro_MOLPRO1,werner2020molpro_MOLPRO2,molpro_website,busch1991analytical_MOLPRO_Gradients}.

The kink in the surfaces around 1.7\AA$ $ is an artifact of the state-averaging, which is likely a result of the significant increase in the $\sigma^*$ orbital energy.
As we are only investigating the ring opening past $2.35$\AA$ $ in this work, we are not concerned by this artifact.

\begin{table}[hptb]
    \centering
    \begin{tabular}{lrrr}
\hline
\hline
\multicolumn{4}{c}{All units are Angstrom} \\
\hline
 & \multicolumn{1}{c}{x} &\multicolumn{1}{c}{y} &\multicolumn{1}{c}{z} \\
\hline
O2&  -1.2375683086   &     0.0000000000  &     -0.2866402151 \\
C1&   0.0602693949   &     0.0000000000  &      0.8872921916 \\
O1&  -0.1338188863   &     0.0000000000  &      2.0472856335 \\
C2&  -0.5677977969   &     0.0000000000  &     -1.5101820972 \\
H&  -0.8589569916   &     0.8787615684  &     -2.0782345615 \\
H&  -0.8589569916   &    -0.8787615684  &     -2.0782345615 \\
C3&   1.2421213373   &     0.0000000000  &      0.0303118387 \\
H&   2.2286743161   &     0.0000000000  &      0.4485450894 \\
C4&   0.9136273524   &     0.0000000000  &     -1.2671537403 \\
H&   1.6169441043   &     0.0000000000  &     -2.0781659814 \\
\hline
    \end{tabular}
    \caption{Furanone geometry at $r\text{C(=O)-O}=1.75$\AA.  This geometry is the starting point for forward and backward scans of slices of the potential energy surfaces.}
    \label{tab:SI:middlegeom}
\end{table}

\begin{figure}[hptb]
    \centering
    \includegraphics[width=\textwidth]{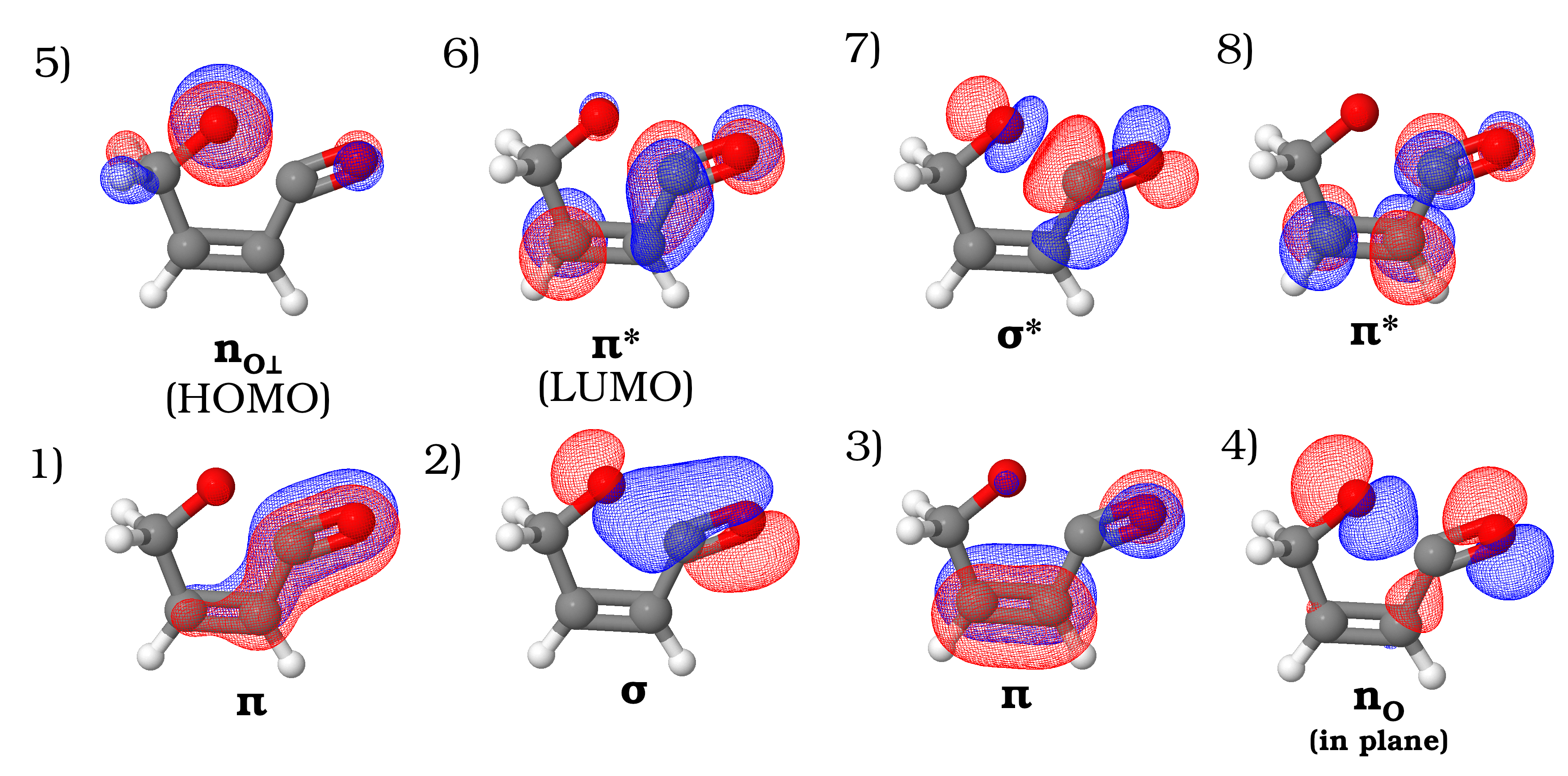}
    \caption{
        Orbitals at the $r\text{C(=O)-O}=1.75$\angs geometry used for the SA-CASSCF surface scans.
        Orbitals 1-5 are occupied in the Aufbau configuration.
        All orbitals are plotted with isosurface value 0.05 utilizing Jmol\cite{jmol}.
    }
    \label{fig:SI:SACASORBS}
\end{figure}

\pagebreak
\section{Geometries studied}
Tables \ref{tab:SI:235Geom}-\ref{tab:SI:345Geom} contain the XYZ geometries of the four furanone ring opening geometries for which X-ray absorption spectra are calculated.
\begin{table}[hptb]
    \centering
    \begin{tabular}{lrrr}
\hline
\hline
\multicolumn{4}{c}{All units are Angstrom} \\
\hline
 & \multicolumn{1}{c}{x} &\multicolumn{1}{c}{y} &\multicolumn{1}{c}{z} \\
\hline
O2&       -1.3239160833  &      0.0000000000  &     -0.9040294567 \\
C1&       -0.0747082902  &      0.0000000000  &      1.0864428214 \\
O1&       -0.4002830183  &      0.0000000000  &      2.2109126967 \\
C2&       -0.2109048071  &      0.0000000000  &     -1.7267662939 \\
H&        -0.2398125597  &      0.8795800823  &     -2.3672661167 \\
H&        -0.2398125597  &     -0.8795800823  &     -2.3672661167 \\
C3&        1.1764225688  &      0.0000000000  &      0.3476401489 \\
H&         2.1052133016  &      0.0000000000  &      0.8883850706 \\
C4&        1.1022767144  &      0.0000000000  &     -0.9916340519 \\
H&         1.9928741547  &      0.0000000000  &     -1.5940692327 \\
\hline
    \end{tabular}
    \caption{Furanone geometry at $r\text{C(=O)-O}=2.35$\angs.}
    \label{tab:SI:235Geom}
\end{table}

\begin{table}[hptb]
    \centering
    \begin{tabular}{lrrr}
\hline
\hline
\multicolumn{4}{c}{All units are Angstrom} \\
\hline
 & \multicolumn{1}{c}{x} &\multicolumn{1}{c}{y} &\multicolumn{1}{c}{z} \\
\hline
O2&       -1.2757257054  &      0.0000000000   &    -1.3203176079 \\
C1&       -0.2260307472  &      0.0000000000   &     1.2214613940 \\
O1&       -0.4556307841  &      0.0000000000   &     2.3786086446 \\
C2&        0.0020150107  &      0.0000000000   &    -1.8562661172 \\
H&         0.1265706382  &      0.8807154155   &    -2.4820251718 \\
H&         0.1265706382  &     -0.8807154155   &    -2.4820251718 \\
C3&        1.0490469094  &      0.0000000000   &     0.4998103445 \\
H&         1.9496215885  &      0.0000000000   &     1.0896657892 \\
C4&        1.1204886195  &      0.0000000000   &    -0.8408543972 \\
H&         2.0961307116  &      0.0000000000   &    -1.2956671993 \\

\hline
    \end{tabular}
    \caption{Furanone geometry at $r\text{C(=O)-O}=2.75$\angs.}
    \label{tab:SI:275Geom}
\end{table}

\begin{table}[hptb]
    \centering
    \begin{tabular}{lrrr}
\hline
\hline
\multicolumn{4}{c}{All units are Angstrom} \\
\hline
 & \multicolumn{1}{c}{x} &\multicolumn{1}{c}{y} &\multicolumn{1}{c}{z} \\
\hline
O2&      -1.2129932845  &      0.0000000000 &      -1.6667248669 \\
C1&      -0.3322203357  &      0.0000000000 &       1.3576327529 \\
O1&      -0.4502044549  &      0.0000000000 &       2.5357455604 \\
C2&       0.1462010534  &      0.0000000000 &      -1.9403850231 \\
H&        0.3884905791  &      0.8802769971 &      -2.5304833664 \\
H&        0.3884905791  &     -0.8802769971 &      -2.5304833664 \\
C3&       0.9265185704  &      0.0000000000 &       0.5893519468 \\
H&        1.8203114412  &      0.0000000000 &       1.1913243743 \\
C4&       1.0806295219  &      0.0000000000 &      -0.7478615493 \\
H&        2.1019829273  &      0.0000000000 &      -1.0914796636 \\

\hline
    \end{tabular}
    \caption{Furanone geometry at  $r\text{C(=O)-O}=3.15$\angs.}
    \label{tab:SI:315Geom}
\end{table}

\begin{table}[hptb]
    \centering
    \begin{tabular}{lrrr}
\hline
\hline
\multicolumn{4}{c}{All units are Angstrom} \\
\hline
 & \multicolumn{1}{c}{x} &\multicolumn{1}{c}{y} &\multicolumn{1}{c}{z} \\
\hline
O2&        -1.1617420053  &      0.0000000000  &     -1.9058915855 \\
C1&        -0.3874493537  &      0.0000000000  &      1.4560975414 \\
O1&        -0.4401034974  &      0.0000000000  &      2.6398450585 \\
C2&         0.2249900729  &      0.0000000000  &     -1.9795772498 \\
H&          0.5425003015  &      0.8793248110  &     -2.5340122495 \\
H&          0.5425003015  &     -0.8793248110  &     -2.5340122495 \\
C3&         0.8486618380  &      0.0000000000  &      0.6427268657 \\
H&          1.7459253857  &      0.0000000000  &      1.2403598879 \\
C4&         1.0357420137  &      0.0000000000  &     -0.6937649489 \\
H&          2.0764005338  &      0.0000000000  &     -0.9764725408 \\

\hline
    \end{tabular}
    \caption{Furanone geometry at $r\text{C(=O)-O}=3.45$\angs.}
    \label{tab:SI:345Geom}
\end{table}

\pagebreak
\section{Excited-state-specific electronic structure}
As mentioned in the main text, our approach for calculating the core-excited states of interest is an excited-state-specific electronic structure method.
Here, we describe in detail the complete procedure for generating the calculated XAS spectra of the main text (Figure 3).

Each of the single geometry calculated XAS spectra contain 16 total transitions.
At a given geometry, for each of the four unique carbon atoms, core-to-SOMO 1, core-to-SOMO 2, and the two spin-coupled core-to-LUMO singlets are calculated.
No excitations to higher energy orbitals are considered.
Each individual excitation energy is calculated via a four stage procedure: a closed-core reference characterization, an excited-state-specific minimal configuration state function (CSF) orbital optimization, a selected configuration interaction (sCI) weak correlation recovery and quasi-rediagonalization, and finally oscillator strength calculation.
We will outline the details of each of these stages in the following subsections.

In all calculations the carbon 1s of interest utilizes the aug-cc-pCVTZ basis set.
For convenience, all other heavy atoms (3 other carbon atoms and 2 oxygen atoms) utilize the pseudopotential ccECP-aug-cc-pVTZ basis, while all hydrogens utilize ccECP-cc-pvtz\cite{bennett2017new_ccECPBasisSets}.
No relativistic effects are accounted for in any calculations presented as these effects are expected to be small at the carbon K-edge\cite{takahashi2017relativistic_RelativisticEffects}.

\subsection{Closed-core reference characterization}
Given a geometry and carbon core of interest, we begin with a restricted Hartree-Fock calculation.
The MP2 natural orbitals are formed by separately diagonalizing the occupied-occupied and virtual-virtual blocks of the MP2-1RDM (i.e., no occupied-virtual mixing is included). 
A copy of these MP2 natural orbitals are saved for later (see the section on weak correlation recovery).
We then perform a 2-state, state-averaged (10e,8o) CASSCF calculation utilizing the MP2 orbitals indexed (12,14,15,16,17,18,19,20), chosen to qualitatively match the active space of the PES scans.
These 2-state, state-averaged CASSCF orbitals provide our initial guess for a set of orbitals, importantly, the approximate shapes for the two SOMOs and $\pi^*$ LUMO--for the next stage of calculation.

\subsection{Excited-state-specific minimal CSF orbital optimization}
The core of our methodology is an excited-state-specific minimal CSF orbital optimization.
This provides a minimal reference mean field set of optimal core-hole relaxed orbitals for all states of interest.
This is done in a CASSCF inspired two step fashion--that is iteratively solving a CI problem followed by optimizing the orbitals for the CI root of interest.
The orbital relaxation is accomplished utilizing previously developed excited state specific state tracking methodology\cite{tran2019tracking_Lan_SSCASSCF} to follow the target root of interest and provide complete orbital relaxation about the core hole.
The CI solver diagonalizes the Hamiltonian in the subspace of $\left<\hat{\text{S}}_\text{z}=0\right>$ determinants with occupations matching the target state of interest.
For the furanone states of interest, the active space is always composed of 4 electrons and 4 orbitals: in order, the carbon 1s of interest, the oxygen $n_\perp$ SOMO1, carbon 1 $\sigma^*$ SOMO2, and the $\pi^*$ LUMO.
A strict restricted active space (RAS) is applied to enforce the orbital occupations.
Importantly, these calculations are entirely state-specific--i.e. there is no state-averaging between different RAS subspaces as each state is a standalone calculation.
The two spin-coupled core-to-LUMO transitions rely upon tracking metrics\cite{tran2019tracking_Lan_SSCASSCF} such as the difference in density matrix to optimize each component individually.

For an example of the RAS employed, refilling the oxygen $n_\perp$ SOMO1 utilizes a RAS of (1,2,1,0)--that is one electron in the 1s, 2 electrons in the oxygen $n_\perp$ SOMO1, 1 electron in the $\sigma^*$ SOMO2, and no electrons in the $\pi^*$ LUMO.
This results in the CI solver determining the single CSF that trivially singlet couples the two unpaired electrons residing in the core and $\sigma^*$ SOMO2 orbitals.
For a core-to-LUMO excitation, the RAS is of the form (1,1,1,1), dictating all four active space orbitals singly occupied.
Solving this Hamiltonian results in a 2-CSF eigenstate--that is the final Hamiltonian eigenstate is a linear combination of two $<S_z>=<S^2>=0$ CSFs, which can be formed via Clebsch-Gordon coefficients.
One set of these Clebsch-Gordon coefficients is given in the further discussion of the Spin Hamiltonian below.

Practically, this stage is accomplished via interfacing DICE\cite{sharma2017semistochastic_DICE_1,holmes2016heat_DICE2,smith2017cheap_DICE_Pyscf,holmes2017excited_DICE_ExcitedStates} and PySCF\cite{sun2018pyscf_PySCF,sun2020recent_PySCF}, where DICE solves the CI problem and PySCF is used to relax the orbitals.

\subsection{Selected CI weak correlation recovery}
The orbitals obtained from the state-specific minimal CSF orbital optimization are the optimal minimal reference, mean field orbitals for each individual state.
We aim to provide some weak correlation energy recovery, as well as to allow the states to remix with one another via selected CI.
Formally an approximation to the full CI wavefunction, selected CI maintains the same combinatorial scaling as full CI.
As such, correlating all electrons requires truncating the orbital space to some practical number of orbitals.
However, any truncation of the full CI wavefunction means the final wavefunction is no longer invariant to the choice of correlating basis.
Further, by performing the state-specific orbital optimization for a minimal CSF reference, the virtual orbitals lose much of their physical meaning as their shapes do not influence the energy expression involved in the orbital optimization (on visual inspection, these virtual orbitals are quite scrambled and difficult to interpret).

To address this issue, we turn to the core filled MP2 natural orbitals from before.
The MP2 natural orbitals have long been recognized as a good starting basis for higher correlation methods\cite{davidson1972properties_NOUses,jensen1988second}.
We replace all virtual orbitals attained from the minimal CSF orbital optimization stage with the MP2 natural orbitals of the same index.
We Gram-Schmidt orthonormalize these orbitals against the minimal CSF orbitals, starting from the first orbital indexed above the orbitals kept from the minimal CSF stage.
One further complication arises--these are ground state orbitals which have not been optimized in the presence of a core-hole.
To address this issue, our final sCI stage is done in two sub-stages.
First, a very large subspace of orbitals is chosen and a selected CI calculation is loosely converged in this subspace.
We then transform our basis into the natural orbitals of the sCI root that most closely matches the single CSF orbital optimized state, determined by a simple non-orthogonal CI (NOCI) overlap of the 25 most important determinants in each sCI root against the minimal CSF state.
This natural orbital transformation includes all occupied-occupied, occupied-virtual, and virtual-virtual rotations.
We then truncate this rotated subspace even further to perform our final sCI stage in the valence subspace of the molecule.
This final, more aggressive sCI wavefunction is the final wavefunction we will use for energy and oscillator strength calculations.
In all core-excited state sCI calculations, the only RAS restriction enforced is that the core 1s of interest only contains exactly 1 electron in all included determinants.

For furanone, 19 orbitals are retained from the minimal CSF orbital optimization stage, while all higher indexed orbitals are replaced with the MP2 natural orbitals. 
We truncate the correlating basis at the number of orbitals that would be contained in the n=3 subshells for all heavy atoms and n=2 subshell for hydrogens (total 34 electrons in 99 orbitals, recalling 5 core orbitals and their electron pairs have been removed with pseudopotential basis sets).
The loosely converged selected CI utilizes a DICE variational expansion selection criteria threshold, $\epsilon_1$, of $\epsilon_1=2.5*10^{-3}$.
For the final correlating basis (ie once we have rotated the subset of 99 orbitals to the natural orbitals of the target root of interest), we truncate the orbital set to include the number of orbitals contained in the n=2 subshell of heavy atoms and n=1 for hydrogens (total 34 electrons in 29 orbitals, again with 5 core orbitals and electron pairs removed by pseudopotential basis sets).
The final selected CI wavefunctions are generated with $\epsilon_1=2*10^{-4}$.
Each time a selected CI calculation is performed, the root of interest is identified via NOCI overlaps of the 25 most important determinants against the minimal CSF reference.

\subsection{Oscillator strengths}
All X-ray oscillator strengths are calculated relative to the core-filled singlet diradical reference.
The core-filled singlet diradical reference state is calculated via the same procedure as the core-excited states, with the only difference being modified restrictions to account for a doubly occupied core 1s orbital.
In the minimal CSF orbital optimization, the active orbitals are the same as in the core-excited case, and the RAS can be written (2,1,1,0), which implies a doubly occupied core orbital, singly occupied SOMO's, and an unoccupied LUMO.
All sCI calculations dictate the core orbital of interest must be doubly occupied in all determinants.

Oscillator strengths are determined by calculating $\left<\Psi_{core\text{-}excited\text{ }state}\right| {\hat{\mu}} \left|\Psi_{diradical}\right>$, where $\hat{\mu}$ contains the three Cartesian components of the dipole operator.
Practically, we project $\hat{\mu}$ into the core-filled diradical reference basis and explicitly act onto the core-filled diradical reference.
We then perform NOCI overlaps between the resulting state and the core-excited state.
For computational efficiency, for both core filled and core-excited states involved in the excitation, we restrict ourselves to the 25 determinants with largest coefficients from the final sCI calculation.
The final oscillator strength is calculated as the standard\cite{hilborn1982einstein} $\frac{2}{3}\Delta E\left|\left<\Psi_{core\text{-}excited\text{ }state}\right| {\hat{\mu}} \left|\Psi_{diradical}\right>\right|^2$, where $\Delta E$ is the energy difference in hartree between diradical and core-excited states (i.e. the excitation energy).
We set the center of molecular charge to the origin\cite{bourne2021reliable_DipoleCalculations}.

\section{Carbon monoxide radical accuracy check}
To check the accuracy of the calculation procedure presented above, we investigate the recently studied carbon monoxide cation XAS\cite{couto2020carbon_COCationXAS} at both the carbon and oxygen edges.
We use the experimentally determined bond length 1.1152\AA $ $ from the CCCBDB\cite{CCCBDB}, and we again use the aug-cc-pCVTZ basis on the core of interest and ccECP-cc-pVTZ basis on the other atom.
All calculations are executed according the procedure above, with a few details altered to handle a core filled doublet single radical reference state.
Only the lowest three states are calculated, which can be described as core-to-SOMO and the two different doublet spin-coupled core-to-LUMO (with 3 unpaired electrons in 3 orbitals) transitions.
The MP2 natural orbitals are taken as that of the closed shell (i.e., neutral CO molecule) at the cationic geometry, and the initial guess orbitals for the minimal CSF calculation are taken from a restricted openshell Hartree Fock calculation for the cation.
The active space is always composed of 4 orbitals, core 1s, SOMO, and both components of the degenerate $\pi^*$ LUMO, although only one of the degenerate excitations is considered.
The large sCI space is composed of 60 orbitals, with $\epsilon_1=5*10^{-3}$.
The final sCI space is composed of 24 orbitals, with $\epsilon_1=2*10^{-4}$.

\begin{figure}[hptb]
    \centering
    \begin{subfigure}[hptb]{\textwidth}
    \centering
        \includegraphics[width=\textwidth]{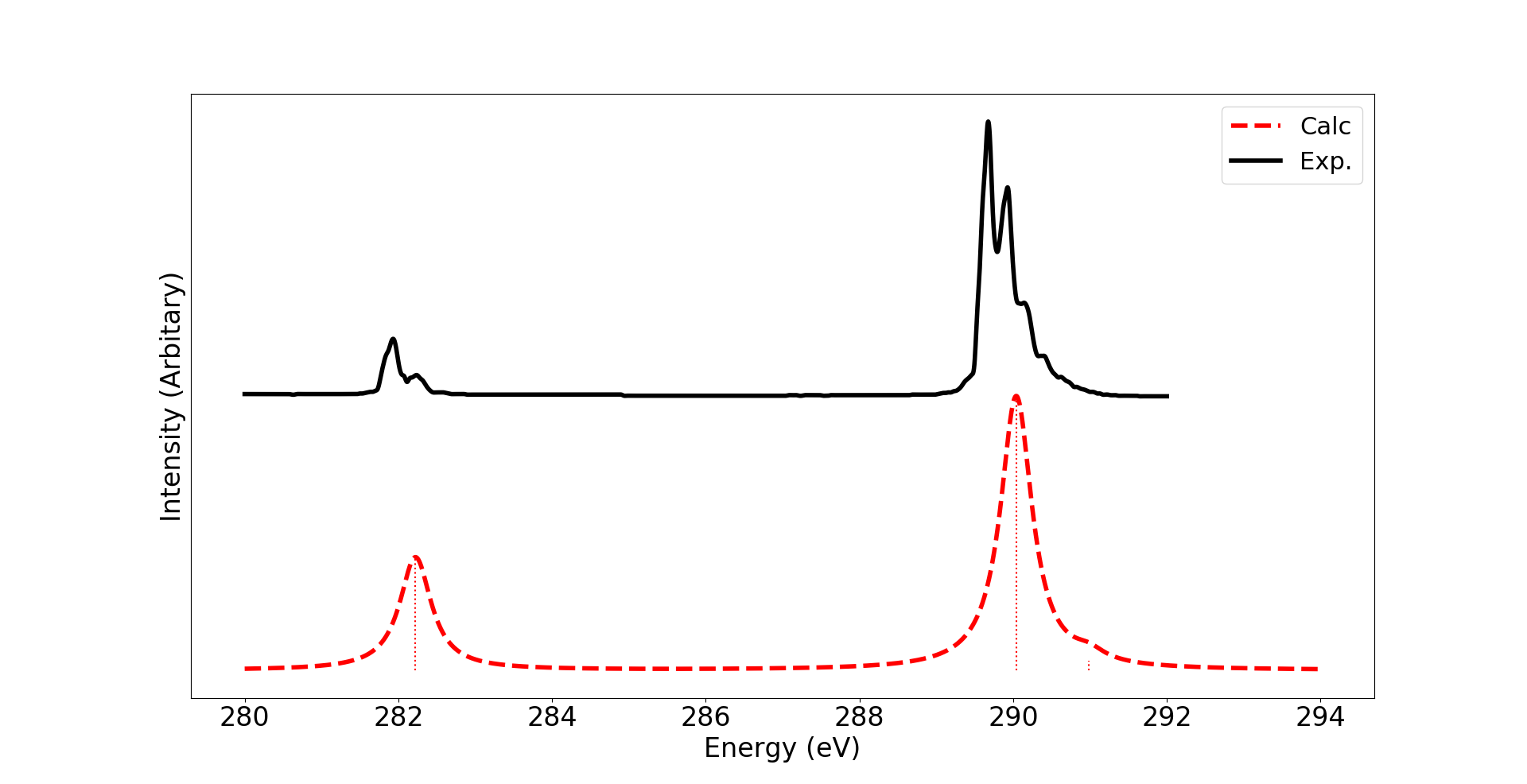}
        \label{fig:SI:CO+_CEdge}        
    \end{subfigure}
    \begin{subfigure}[hptb]{\textwidth}
    \centering
        \includegraphics[width=\textwidth]{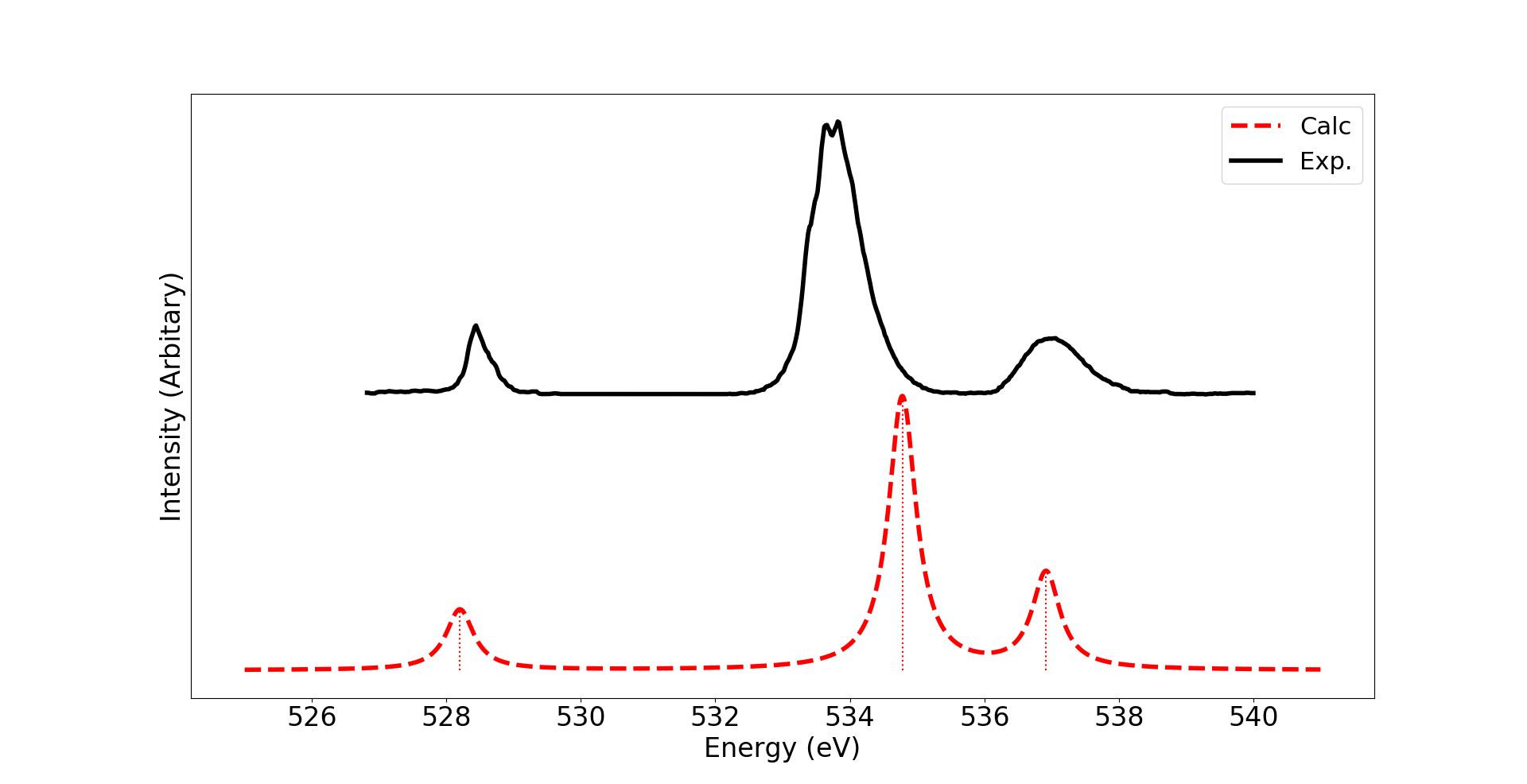}
        \label{fig:SI:CO+_OEdge}        
    \end{subfigure}    
    \caption{Calculated X-Ray absorption spectra for CO$^+$ both carbon (top) and oxygen (bottom) K-edges.
    No empirical shifting of the calculated spectra is employed.
    All states are broadened with Lorentzian functions with width of 0.5 eV.
    The experimental spectrum is digitized from \citet{couto2020carbon_COCationXAS}.
    }
    \label{fig:SI:CO+}
\end{figure}

The results of the calculations are shown in Figure \ref{fig:SI:CO+}.
At the carbon edge (Figure \ref{fig:SI:CO+}, top), the absolute excitation energies, spacing between excitations, and relative oscillator strengths all qualitatively and quantitatively reproduce the experimental spectra. 
The two spin-coupled core-to-LUMO transitions are both shown, but the higher energy component of these states has relatively small oscillator strength and is buried in the vibrational structure of the lower energy component in the experiment.
No vibrational structure is accounted for in the calculated spectra.

For the oxygen edge (Figure \ref{fig:SI:CO+}, bottom), a qualitatively correct picture is obtained.
The absolute excitation energies of the first and third transitions are in agreement with experiment, while the calculation overestimates the excitation energy of the second transition.
This overestimate of energy ensuingly leads to a slight under estimate of the splitting between the two doubled spin-coupled core-to-LUMO states.
The relative intensities of all three transitions are in good agreement with the experimentally observed states.

Overall, we believe this data provides confidence in the accuracy of the electronic structure method.
By explicitly including orbital relaxation in the minimal CSF orbital optimization stage, our calculated absolute excitation energies are in good agreement with experiment.
We emphasize again that no empirical shift is included for the calculated data.
Particularly reassuring to the conclusions of this work is the consistent qualitative agreement of relative oscillator strengths.
For both carbon and oxygen K-edges, the relative intensities of both spin-coupled components of the core-to-LUMO transition, as well as these intensities relative to the refilling core-to-SOMO transition, agree with the experimentally observed spectra.
Because the multi-reference nature of our approach can naturally handle the diradical and core-excited-atop-a-diradical states, we expect the method to be similarly accurate for diradical states.
However to the best of our knowledge, no clear experimental spectra are available to test against.

\section{Furanone Ground State XAS}
The geometry used for the spectra is given in Table \ref{tab:SI:GSGeom}.
The calculations for the ground state spectra only calculate the core-to-LUMO excitation for all unique carbon atoms.
For the aufbau ground state, the minimal CSF orbital optimization is equivalent to Hartree Fock, and all virtual orbitals are replaced by the virtual MP2 natural orbitals for the sCI weak correlation recovery stages.
For some calculations, we found it useful to seed the minimal CSF orbital optimization stage with orbitals attained from a $\Delta$SCF/MOM calculation for a core-to-LUMO excited state.
For these calculations, the aufbau occupied orbital shapes are taken from the $\Delta$SCF/MOM calculation, as they have already been relaxed for the core hole, and the remaining orbitals are taken as the MP2 natural orbitals.
All other components of the calculation procedure remain identical to the procedure outlined above.

The calculated spectra for these is shown in Figure \ref{Fig:SI:RingClosedSpectra}.
The general shape is similar to the reported ground state spectra of furan\cite{severino2022non_furanringopenXAS} or benzene\cite{vidal2020interplay_BenezenPlus_Theory}, with the lowest energy feature being intense core-to-$\pi^*$ LUMO around 286 eV.
As we are neglecting all excitations beyond core-to-LUMO, we do not have additional structure
In both furan\cite{severino2022non_furanringopenXAS} and benzene\cite{vidal2020interplay_BenezenPlus_Theory}, all additional structure in the ground state spectra is attributed to Rydberg excitations, which are generally weaker and lie energetically above the primary core-to-LUMO feature.
This provides confidence that our predictions regarding the ring-open core-to-LUMO spin-coupled peaks around 285.5 eV (main text Figures 3 and 4) will remain in a relatively isolated portion of the spectrum free from overlapping transitions.

\begin{table}[hptb]
    \centering
    \begin{tabular}{lrrr}
\hline
\hline
\multicolumn{4}{c}{All units are Angstrom} \\
\hline
 & \multicolumn{1}{c}{x} &\multicolumn{1}{c}{y} &\multicolumn{1}{c}{z} \\
\hline
O2& -1.0901778534 &       0.0000000000    &   -0.0310840570\\
C1&  0.0477907079 &       0.0000000000    &    0.7844081584\\
O1&  0.0040331194 &       0.0000000000    &    1.9756523786\\
C2& -0.7168265384 &       0.0000000000    &   -1.3935030063\\
H&  -1.1289743311 &       0.8802128201    &   -1.8706862151\\
H&  -1.1289743311 &      -0.8802128201    &   -1.8706862151\\
C3&  1.2139409986 &       0.0000000000    &   -0.1124227004\\
H&   2.2212103496 &       0.0000000000    &    0.2515523651\\
C4&  0.7885400407 &       0.0000000000    &   -1.3850468707\\
H&   1.3876656044 &       0.0000000000    &   -2.2743932506\\
\hline
    \end{tabular}
    \caption{Furanone ground state geometry.}
    \label{tab:SI:GSGeom}
\end{table}

\begin{figure}[hptb]
    \includegraphics[width=\textwidth]{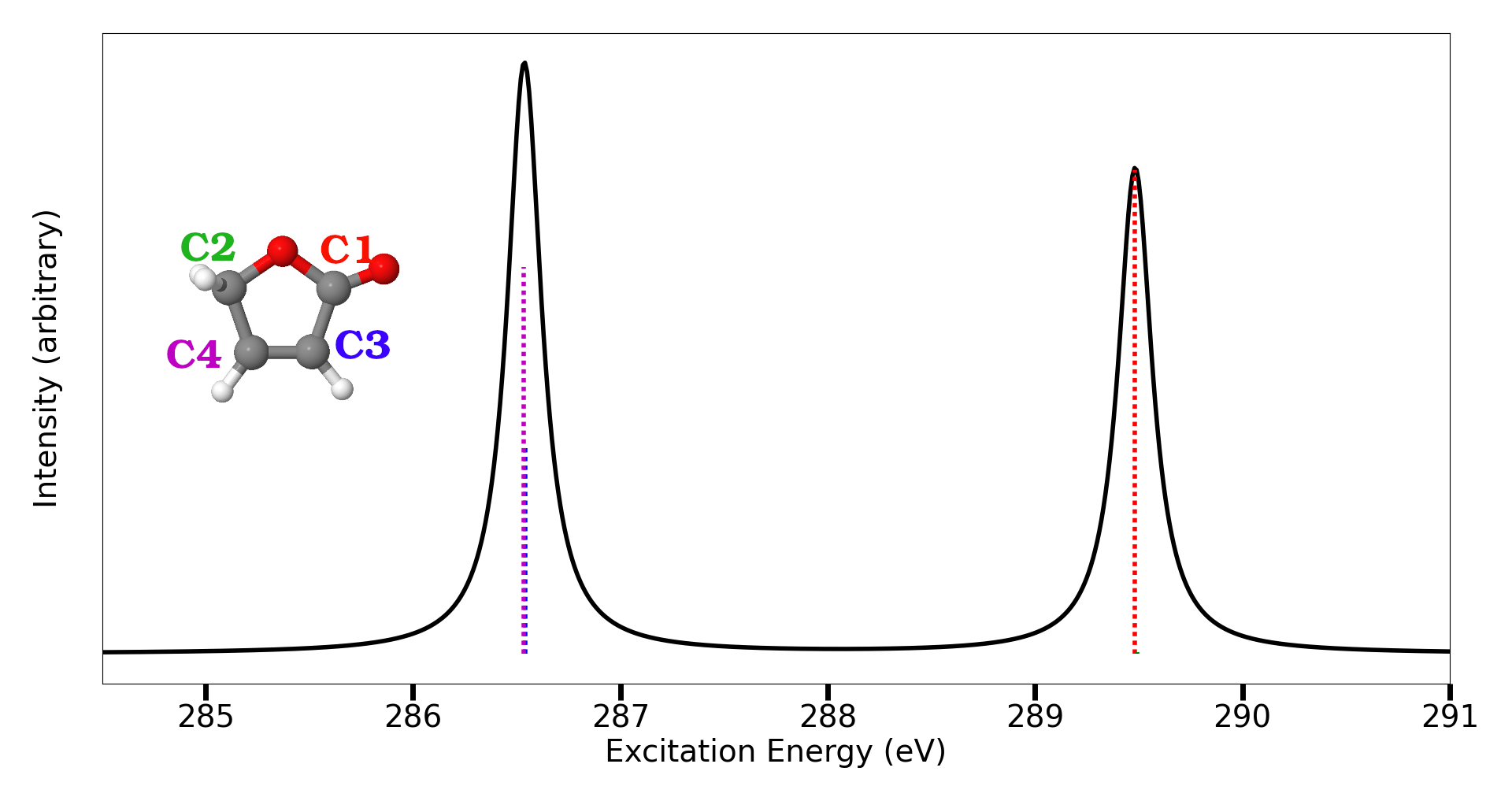}
    \caption{Calculated XAS absorption peaks for core-to-LUMO for all unique carbon atoms.
    The excitation energies from the carbon 3 and 4 core orbitals lie on top of one another at about 285.5 eV.
    The excitation from the carbon 2 core orbital has negligible intensity, but lies at about the same energy as the excitation from the carbon 1 core orbital around 289.5 eV.
    All sticks are broadened with Lortentzian functions with width 0.2 eV.}
    \label{Fig:SI:RingClosedSpectra}
\end{figure}

\pagebreak

\section{Spin Hamiltonian}
\label{sec:spinHamiltonian}
\subsection{Hamiltonian details}
Assuming all couplings are identical and taking the approach of coupling the first two spins, then iteratively coupling the third and fourth spins to the resulting states, utilizing Clebsch-Gordon coefficients gives
\begin{subequations}
\begin{equation}
    \ket{S=0_1} = \frac{1}{\sqrt{12}}\left( 2\eaabb - \eabab - \eabba + 2 \ebbaa - \ebaba - \ebaab \right )
\end{equation}
\begin{equation}
    \ket{S=0_2}=\frac{1}{2}\left(\eabab - \eabba + \ebaba - \ebaab\right)
\end{equation}
\end{subequations}
Intuitively, these linear combinations can be understood as follows.
For $\ket{S=0_1}$, the first and second electrons are triplet coupled ($<S^2>=2$ in both $<S_z>=1$ and $<S_z>=0$ projections, as are the third and fourth electrons.
These triplet coupled pairs of electrons are then coupled together based on the proper pairings of $S_z$ projections to arrive at the $<S^2>=<S_z>=0$ singlet.
For $\ket{S=0_2}$, the first and second electrons are singlet coupled as are the third and fourth, and these subsystems are then coupled together.
These singlets are degenerate under $\hat{S}^2$, so any linear combination of these states remains a single eigenstate.
Therefore these states are not enough to uniquely identify the spectroscopic features nor estimate their oscillator strengths.

It is ultimately the Hamiltonian that determines the linear combination of CSFs which is spectroscopically observable.
The 6x6 Hamiltonian matrix of the Heisenberg spin Hamiltonian (main text equation (1)) in the basis $\{$\aabb,\abab,\abba,\bbaa,\baba,\baab$\}$ can be written as
\begin{equation}
\begin{bmatrix}
\myatop{J_{13}+J_{14}}{+J_{23}+J_{24}} & -J_{23} & -J_{24} & 0 & -J_{14} & -J_{13} \\
-J_{23} & \myatop{J_{12}+J_{14}}{+J_{23}+J_{34}} & -J_{34} & -J_{14} & 0 & -J_{12} \\
-J_{24} & -J_{34} & \myatop{J_{12}+J_{13}}{+J_{24}+J_{34}} & -J_{13} & -J_{12} & 0 \\
0 & -J_{14} & -J_{13} & \myatop{J_{13}+J_{14}}{+J_{23}+J_{24}} & -J_{23} & -J_{24} \\
-J_{14} & 0 & -J_{12} & -J_{23} & \myatop{J_{12}+J_{14}}{+J_{23}+J_{34}} & -J_{34} \\
-J_{13} & -J_{12} & 0 & -J_{24} & -J_{34} & \myatop{J_{12}+J_{13}}{+J_{24}+J_{34}} \\
\end{bmatrix}
\end{equation}

\subsection{Example of exchange couplings}
We provide an example of the exponential decay of exchange couplings with a set of calculated couplings.
The couplings shown in Figure \ref{fig:SI:C1CouplingsExample} are between the carbon 1 core 1s, oxygen $n_\perp$ SOMO, $\sigma^*$ SOMO, and $\pi^*$ LUMO as a function of ring opening geometry for one of the core-to-LUMO spin-coupled states.
\begin{figure}[hptb]
    \centering
    \includegraphics[width=\textwidth]{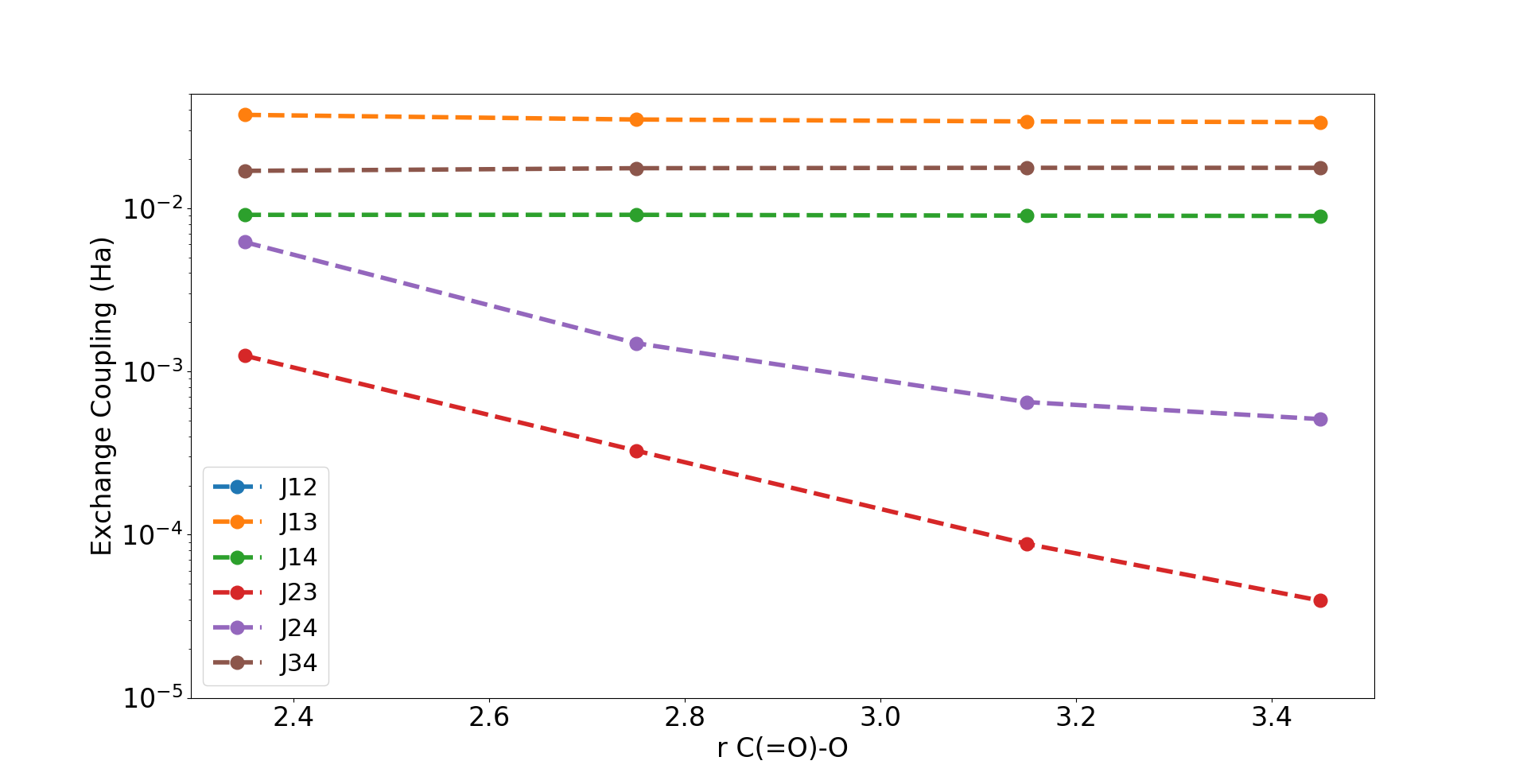}
    \caption{
    Exchange couplings for one of the carbon 1 core-to-LUMO spin-coupled singlets as a function of ring opening coordinate.
    The $i,j$ indicies refer to the core 1s, oxygen $n_\perp$ SOMO, $\sigma^*$ SOMO, and $\pi^*$ LUMO, respectively.
    Notably, the $J_{23}$ coupling--that is, the exchange coupling between the two SOMOs that are spatially being separated as a function of ring opening--is approximately linear on a log scale, indicating exponential decay.
    }
    \label{fig:SI:C1CouplingsExample}
\end{figure}

\pagebreak

\section{Carbon 3}
\subsection{Minimal reference versus sCI}
For carbon 3, the calculated core-to-LUMO spectra (main text Figure 4) shows no qualitative change as a function of ring opening after appropriate broadening of the individual states is applied.
That said, the underlying transitions do change, as indicated by the stick plots.
At $r\text{C(=O)-O}=2.35\text{\AA}$ (main text Figure 4, left), the majority of the intensity is in the higher energy component, while the lower energy component has close to negligible intensity.
As the ring opens, this picture transforms into one where each of the two components has roughly equal intensity (main text Figure 4, left).
This is perhaps surprising considering that the minimal reference spin occupancy compositions of these two states remains relatively unchanged (main text Figure 5) as the ring opens.

We attribute this difference between relatively constant character in the minimal reference state compositions and changing stick heights in the final predicted spectra to the difference between the minimal reference and sCI wavefunctions.
Again, the intensities in the calculated spectra are calculated via the dipole matrix elements of the sCI wavefunctions (that is, the oscillator strengths are proportional to $\left|\left< \Psi_{1s\rightarrow LUMO}^{sCI}\right|\hat{\mu}\left|\Psi_{diradical}^{sCI}\right> \right|^2$, where the superscript $sCI$ indicates these are the sCI wavefunctions).
During the sCI calculation, because the only RAS that is applied is the one where there is a single electron in the 1s of interest, all core-excited states can remix with one another.
This allows the two components of the spin-coupled core-to-LUMO excitation to remix based on the details of a now weakly correlated CI Hamiltonian.
That is, the minimal reference states whose spin occupancy compositions are plotted in the main text Figure 6 can remix via the sCI calculation.

In this regard, we will reuse our 25 determinant NOCI metric (which is primarily intended to identify the sCI root most similar to the minimal reference root) to estimate the amount of remixing between the two spin-coupled core-to-LUMO states for carbon 3.
In addition, we have calculated $\left|\left< \Psi_{1s\rightarrow LUMO}^{min-reference}\right|\hat{\mu}\left|\Psi_{diradical}^{min-reference}\right> \right|^2$, where the wavefunctions are taken from the minimal reference orbital optimization stage for both diradical and core-excited states.
These results are summarized in Table \ref{tab:SI:c3_calcs}.

\begin{table}[hptb]
    \centering
    \begin{tabular}{l|c  c c}
    \hline
    \multicolumn{4}{c}{$r\text{C(=O)-O}=2.35\text{\AA}$} \\
    \hline
    Component & $\left|\left< \Psi_{1s\rightarrow LUMO}^{min-reference}\right|\hat{\mu}\left|\Psi_{diradical}^{min-reference}\right> \right|^2$ & $\left|\left< \Psi_{1s\rightarrow LUMO}^{sCI}\right|\hat{\mu}\left|\Psi_{diradical}^{sCI}\right> \right|^2$ &  NOCI \\
    \hline
    Higher Energy & $3.37*10^{-3}$  & $4.07*10^{-3}$ & 0.67 \\
    Lower Energy & $2.66*10^{-3}$  & $5.94*10^{-5}$ & 0.71  \\   
    \hline
     \hline
    \multicolumn{4}{c}{$r\text{C(=O)-O}=3.45\text{\AA}$} \\
    \hline
    Component & $\left|\left< \Psi_{1s\rightarrow LUMO}^{min-reference}\right|\hat{\mu}\left|\Psi_{diradical}^{min-reference}\right> \right|^2$ & $\left|\left< \Psi_{1s\rightarrow LUMO}^{sCI}\right|\hat{\mu}\left|\Psi_{diradical}^{sCI}\right> \right|^2$ &  NOCI \\
    \hline
    Higher Energy &  $2.54*10^{-3}$ & $2.39*10^{-3}$ & 0.90 \\
    Lower Energy &  $3.52*10^{-3}$ &$2.92*10^{-3}$ & 0.93  \\     
    \hline
         
    \end{tabular}
    \caption{Estimates of the transition dipole moment matrix elements for the two spin-coupled core-to-LUMO singlets for carbon 3 in the furanone ring opening, as well as a NOCI estimate of the overlap of the sCI and minimal reference wavefunctions.
    The NOCI estimate approximates $\left|\left<\Psi_{1s\rightarrow LUMO} \right | \left . \Psi_{1s\rightarrow LUMO}^{sCI}\right>\right|$.
    Lower (Upper) refers to the energy ordering of the lower (higher) energy spin-coupled singlet at a given geometry.
    }
    \label{tab:SI:c3_calcs}
\end{table}

Starting with the transition dipole moment estimate from the sCI wavefunctions, 
(that is, $\left|\left< \Psi_{1s\rightarrow LUMO}^{sCI}\right|\hat{\mu}\left|\Psi_{diradical}^{sCI}\right> \right|^2$), at $r\text{C(=O)-O}=2.35\text{\AA}$, the higher energy component is two orders of magnitude larger than the lower energy component -- indeed these are the matrix elements that primarily determine the relative intensities of the blue sticks in the main text Figure 4, left.
At $r\text{C(=O)-O}=3.45\text{\AA}$, both spin-coupled component matrix elements have the same order of magnitude estimate, which again is what determines the approximately equivalent height (blue sticks in the main text Figure 4, left).
However, when the transition dipole moment is estimated from the minimal reference wavefunctions, at either geometry, both matrix elements are on the same order of magnitude.
Further, at both geometries, the magnitude of the estimate remains constant. 
This is consistent with the understanding derived from inspecting the spin occupancies presented in the main text Figure 5--no matter the geometry, both components are dominated by dipole-allowed spin occupancies and therefore should have appreciable intensity. 

Inspecting the NOCI overlaps, one can attribute this difference in qualitative description to the remixing of states in the sCI calculation.
At $r\text{C(=O)-O}=2.35\text{\AA}$, the singlets remix amongst themselves strongly, so much so that the overlap of the lower energy component sCI state with its minimal reference equivalent is 0.71 (and in fact, the NOCI overlap of this sCI root with the other minimal reference spin-coupled component in the same orbital basis is 0.60).
Critically, this implies that the introduction of weak correlation and quasi-rediagonalization with other nearby states strongly remixes the minimal reference states at this geometry.
This remixing creates a scenario where even though the minimal reference states are both expected to have appreciable intensity (because of significant composition of dipole allowed spin occupancies), the weakly correlated sCI states put the majority of this intensity in one of the two components.
At $r\text{C(=O)-O}=3.45\text{\AA}$, the large NOCI overlap metrics ($\geq$0.9) indicate that there is significantly less remixing between the minimal reference states via the sCI calculation.
Due to this lack of remixing, the minimal reference and sCI matrix elements are in agreement that both states should have appreciable intensity, which is consistent with the spectra (Figure 4) and minimal reference analysis (Figure 5) of the main text.

It is important to reiterate that while this analysis identifies the differences between the minimal reference and sCI wavefunction's matrix elements, we qualitatively predict very little time dependent signal for this transition once the appropriate broadening is applied.
This is a result of the near degeneracy of the two transitions (the spin splitting never exceeds 0.2 eV).
One might expect that because the states are so close energetically they always remix strongly when going from minimal reference to sCI wavefunctions; however the fact that the mixing is minimal at the ring open geometry emphasizes that this remixing depends strongly on the details of weak correlation and therefore cannot be predicted \textit{a priori}.

Even with this strong dependence on the details of the high-level calculation, we apply our simple H\"{u}ckel-like approximations to the carbon 3 core-to-LUMO XAS states.
The couplings for this carbon are shown in Figure \ref{fig:SI:C3SquareExchange} and the results of diagonalizing the spin Hamiltonian are shown in Table \ref{tab:SI:c3_huckel}.
In this case, the most interesting result from the H\"{u}ckel-like approximations to the exchange couplings is that at the ring open geometry ($r\text{C(=O)-O}=3.45\text{\AA}$) the spin Hamiltonian eigenstates are degenerate.
This degeneracy prediction is of course a result of the aggressively simple approximation made when applying the H\"{u}ckel-like approximations to the spin coupling, as the high-level calculation indeed predicts some splitting between these states.

\begin{figure}[hptb]
    \centering
    \begin{subfigure}{0.45\textwidth}
        \centering
        \includegraphics[width=\textwidth]{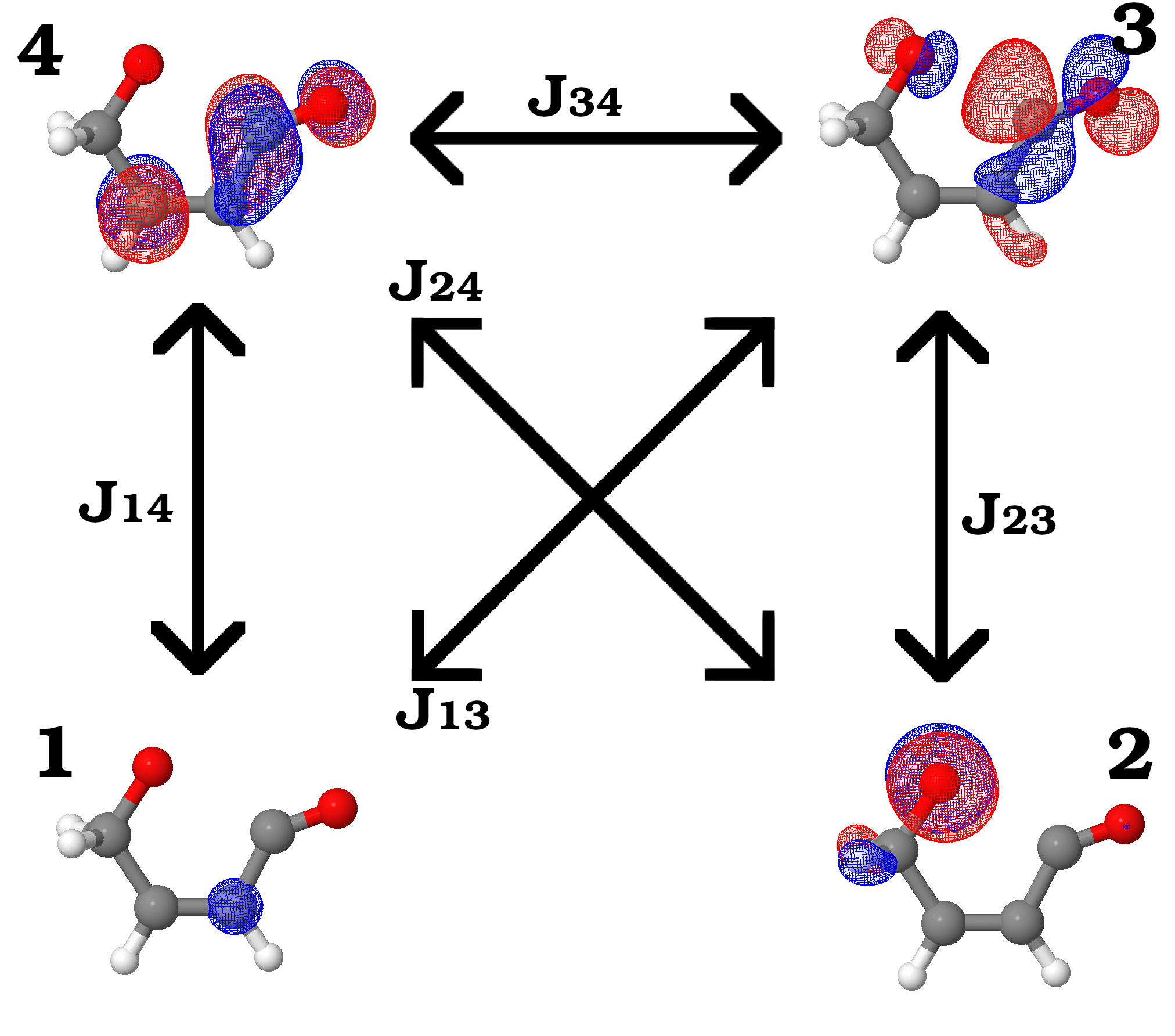} 
        \caption{$r\text{C(=O)-O} = 2.35\text{\AA}$}
    \end{subfigure}\hfill
    \begin{subfigure}{0.45\textwidth}
        \centering
        \includegraphics[width=\textwidth]{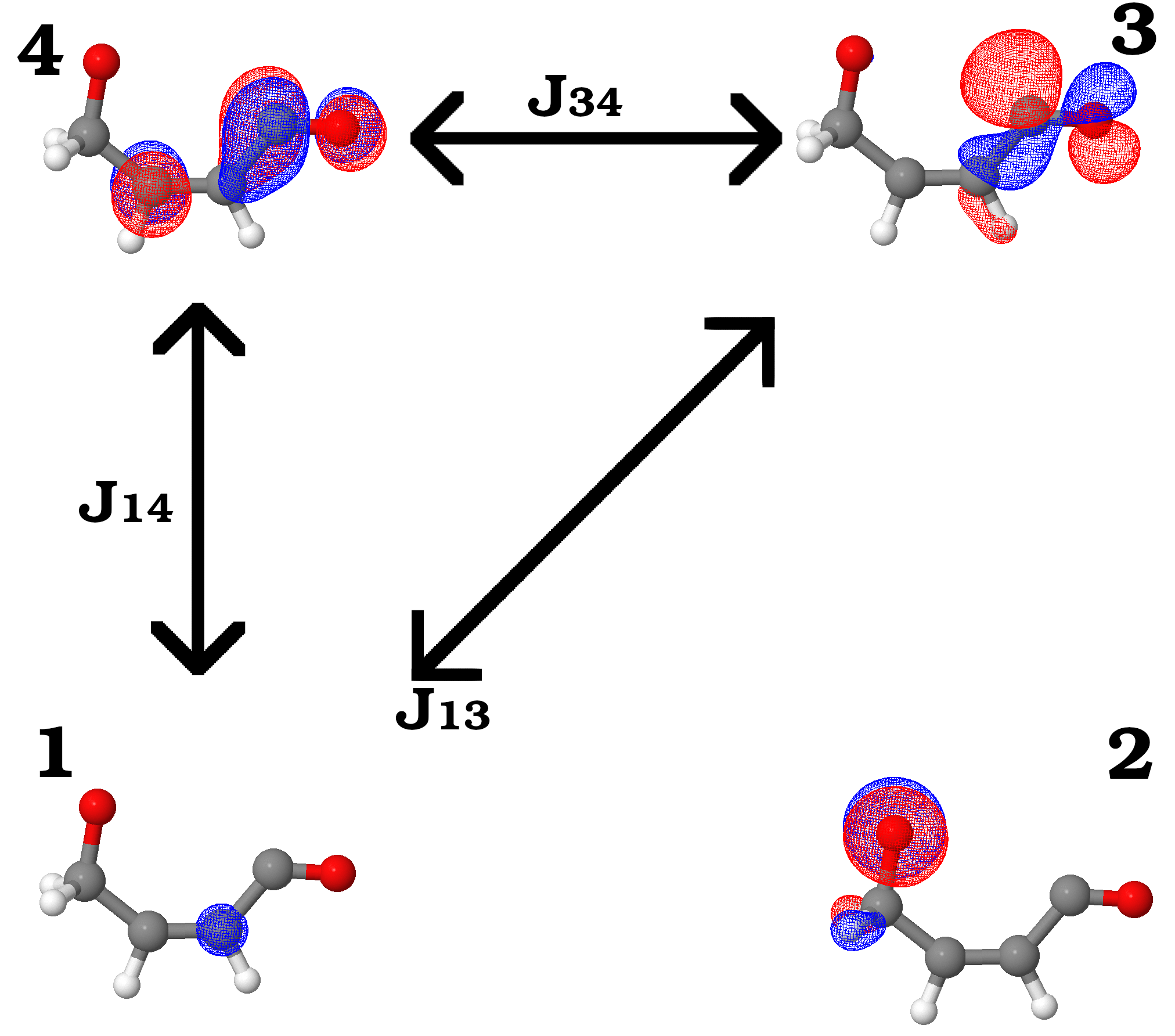} 
        \caption{$r\text{C(=O)-O} = 3.45\text{\AA}$}    
    \end{subfigure}
    \caption{Exchange couplings for carbon 3 in the furanone photochemical ring opening.  
    The orbitals ordering and the interpretation of ``on" versus ``off" couplings is identical to that in Figure 7 of the main text.
    }
    \label{fig:SI:C3SquareExchange}
\end{figure}

\begin{table}[hptb]
    \centering
    \begin{tabular}{l |c|c}
         \hline
         \multicolumn{3}{c}{$r\text{C(=O)-O}=2.35\text{\AA}$} \\
         \hline
         \multicolumn{3}{c}{$J_{13}=J_{14}=J_{23}=J_{24}=J_{34}\neq0$, $\quad$$J_{12} = 0$}\\
         \hline
         Component & $c_{bright}$ & $c_{dark}$ \\
         \hline
         Higher Energy & 0.83 & 0.17 \\
         Lower Energy & 0.50 & 0.50 \\
         \hline
         \hline
         \multicolumn{3}{c}{$r\text{C(=O)-O}=3.45\text{\AA}$} \\
         \hline
         \multicolumn{3}{c}{$J_{13}=J_{14}=J_{34}\neq0$, $\quad$$J_{12}=J_{23}=J_{24} = 0$}\\
         \hline
         Component & $c_{bright}$ & $c_{dark}$ \\
         \hline
         Higher Energy & 0.83$^*$ & 0.17$^*$ \\
         Lower Energy & 0.50$^*$ & 0.50$^*$ \\
         \hline
    \end{tabular}
    \caption{
    Relative weights of X-ray optically allowed versus disallowed spin occupancies of the spin Hamiltonian singlet eigenstates for carbon 3 core-to-LUMO transitions formed by estimating couplings with H\"{u}ckel-like approximations.
    At $r\text{C(=O)-O}=3.45\text{\AA}$, the two singlet eigenstates of the spin Hamiltonian are degenerate, meaning the two eigenstates found can remix arbitrarily making the coefficients indicated with $^*$ arbitrary.
    The meanings of $c_{bright}$ and $c_{dark}$ are identical to those in Table 1 of the main text.
    }
    \label{tab:SI:c3_huckel}
\end{table}

\section{H\"{u}ckel model implications}
\subsection{Carbons 1 and 2}
Utilizing the H\"{u}ckel-like approximations of exchange coupling, we briefly address the two carbons not analyzed previously--carbons 1 and 2.
For carbon 1, (the C(=O)-O carbon atom), using these H\"{u}ckel-like approximations and solving the spin Hamiltonian results in degenerate singlets at both extremes of the ring opening coordinate (at $r\text{C(=O)-O}=2.35\text{\AA}$ all couplings are considered ``on" while at $r\text{C(=O)-O}=3.45\text{\AA}$ the couplings are identical to those in Figure \ref{fig:SI:C3SquareExchange} (b)).
The calculated spin splitting between the core-to-LUMO singlets is always less than 0.23 eV; when plotting with the appropriate broadening it can be expected that these states always appear as a single feature independent of geometry (indeed this is the red dashed line feature at about 287 eV in the calculated spectra in the main text Figure 3 a)).
For carbon 2, it can be predicted that the core-to-LUMO excitations will have negligible intensity (at least compared to the other carbon atoms) because this carbon does not contribute to the $\pi$ system and therefore the dipole matrix element between the core 1s and $\pi^*$ orbitals is likely negligible.
From our H\"{u}ckel-like perspective, even though the 1s to $\pi^*$ might be considered ``on" from nearest neighbor arguments, one can imagine augmenting the model with a constraint that the two orbitals involved in the primary excitation must contain on atom overlap--which is similar to the local selection rules typically invoked for core-to-valence excitations.

Even though for carbons 1, 2 and 3 this simple H\"{u}ckel-like model does not create such a clear geometry dependent picture as it did for carbon 4, we believe there is potentially still use in a simple model in this situation.
Having predicted minimal spin splitting via degenerate singlets from the spin Hamiltonian may assist both theory and experiment in approximating which atoms may show significant spin-split core-to-LUMO signatures and which may be unresolvable due to broadening.
Further, quickly applying this model across a geometry coordinate may assist in identifying interesting geometry dependent features--as we found for the carbon 4 atom in the furanone ring opening.
However, without experimental or additional theoretical evidence--and in both cases extended to edges beyond the carbon K-edge--making such simplifying H\"{u}ckel-like approximations remains to be validated.

\end{suppinfo}

\bibliography{main}

\end{document}